\documentclass[english]{IEEEtran}
\usepackage[T1]{fontenc}
\usepackage[latin9]{inputenc}
\usepackage{color}
\usepackage{amsthm}
\usepackage{amsmath}
\usepackage{amssymb}
\usepackage{graphicx}

\makeatletter

\newcommand{\lyxdot}{.}

\theoremstyle{plain}
\newtheorem{thm}{\protect\theoremname}
\theoremstyle{plain}
\newtheorem{prop}[thm]{\protect\propositionname}
\theoremstyle{plain}
\newtheorem{lem}[thm]{\protect\lemmaname}

\usepackage{algorithm}
\usepackage{algpseudocode}

\@ifundefined{showcaptionsetup}{}{%
 \PassOptionsToPackage{caption=false}{subfig}}
\usepackage{subfig}
\makeatother

\usepackage{babel}
\providecommand{\lemmaname}{Lemma}
\providecommand{\propositionname}{Proposition}
\providecommand{\theoremname}{Theorem}

\begin{document}

\title{Robust Estimation of Structured Covariance Matrix for Heavy-Tailed
Elliptical Distributions}

\author{Ying~Sun,~Prabhu~Babu,~and~Daniel~P.~Palomar,~\IEEEmembership{Fellow,~IEEE}%
\thanks{Ying~Sun,~Prabhu~Babu,~and~Daniel~P.~Palomar are with the Hong
Kong University of Science and Technology (HKUST), Hong Kong. E-mail:
\{ysunac,~eeprabhubabu,~palomar\}@ust.hk.%
}%
\thanks{This work was supported by the Hong Kong RGC 16207814 research grant.%
}%
\thanks{Part of the results in this paper were preliminary presented at \cite{Sun2015Robust}.%
}}
\maketitle
\begin{abstract}
This paper considers the problem of robustly estimating a structured
covariance matrix with an elliptical underlying distribution with
known mean. In applications where the covariance matrix naturally
possesses a certain structure, taking the prior structure information
into account in the estimation procedure is beneficial to improve
the estimation accuracy. We propose incorporating the prior structure
information into Tyler's \textsl{M}-estimator and formulate the problem
as minimizing the cost function of Tyler's estimator under the prior
structural constraint. First, the estimation under a general convex
structural constraint is introduced with an efficient algorithm for
finding the estimator derived based on the majorization minimization
(MM) algorithm framework. Then, the algorithm is tailored to several
special structures that enjoy a wide range of applications in signal
processing related fields, namely, sum of rank-one matrices, Toeplitz,
and banded Toeplitz structure. In addition, two types of non-convex
structures, i.e., the Kronecker structure and the spiked covariance
structure, are also discussed, where it is shown that simple algorithms
can be derived under the guidelines of MM. Numerical results show
that the proposed estimator achieves a smaller estimation error than
the benchmark estimators at a lower computational cost.\end{abstract}


\section{Introduction}

Estimating the covariance matrix is a ubiquitous problem that arises
in various fields such as signal processing, wireless communication,
bioinformatics, and financial engineering \cite{Ledoit2004,bickel2008regularized,Ollila2012Survey}.
It has been noticed that the covariance matrix in some applications
naturally possesses some special structures. Exploiting the structure
information in the estimation process usually implies a reduction
in the number of parameters to be estimated, and thus is beneficial
to improving the estimation accuracy \cite{fuhrmann1988existence}.
Various types of structures have been studied. For example, the Toeplitz
structure with applications in time series analysis and array signal
processing was considered in \cite{dembo1989embedding,fuhrmann1988existence,miller1987role}.
A sparse graphical model was studied in \cite{friedman2008sparse},
where sparsity was imposed on the inverse of the covariance matrix.
Banding or tapering the sample covariance matrix was proposed in \cite{bickel2008regularized}.
A spiked covariance structure, which is closely related to the problem
of component analysis and subspace estimation, was introduced in \cite{visuri2001subspace}.
Other structures such as group symmetry and the Kronecker structure
were considered in \cite{werner2008estimation,Parikshit2011group,Wirfalt2014kronecker}.

While the previously mentioned works have shown that enforcing a prior
structure on the covariance estimator improves its performance in
many applications, most of them either assume that the samples follow
a Gaussian distribution or attempt to regularize the sample covariance
matrix. It has been realized that the sample covariance matrix, which
turns out to be the maximum likelihood estimator of the covariance
matrix when the samples are assumed to be independent identically
normally distributed, performs poorly in many real-world applications.
A major factor that causes the problem is that the distribution of
a real-world data set is often heavy-tailed or contains outliers.
In this case, a single erroneous observation can lead to a completely
unreliable estimate \cite{zoubir2012robust}.

A way to address the aforementioned problem is to find a robust structured
covariance matrix estimator that performs well even if the underlying
distribution deviates from the Gaussian assumption. One approach is
to refer to the minimax principle and seek the ``best'' estimate
of the covariance for the worst case noise. To be precise, the underlying
probability distribution of the samples $f\left(\cdot\right)$ is
assumed to belong to an uncertainty set of functions $\mathcal{F}$
that contains the Gaussian distribution, and the desired minimax robust
estimator is the one whose maximum asymptotic variance over the set
$\mathcal{F}$ is less than that of any other estimator. Two types
of uncertainty sets $\mathcal{F}$, namely the $\varepsilon$-contamination
and the Kolmogorov class, were considered in \cite{williams1993robust},
where a structured maximum likelihood type estimate (\textsl{M}-estimate)
was derived as the solution of a constrained optimization problem.
The uncertainty set that we are interested in is the family of elliptically
symmetric distributions. It was proved by Tyler in \cite{tyler1987distribution}
that given $K$-dimensional independent and identically distributed
(i.i.d.) samples $\left\{ {\bf x}_{i}\right\} _{i=1,\ldots,N}$ drawn
from an elliptical distribution, the Tyler's estimator defined as
the solution to the fixed-point equation
\[
{\bf R}=\frac{K}{N}\sum_{i=1}^{N}\frac{{\bf x}_{i}{\bf x}_{i}^{T}}{{\bf x}_{i}^{T}{\bf R}^{-1}{\bf x}_{i}},
\]
is a minimax robust estimator. Additionally, it is ``distribution-free''
in the sense that its asymptotic variance does not depend on the parametric
form of the underlying distribution.

The problem of obtaining a structured Tyler's estimator was investigated
in the recent works \cite{soloveychik2013group} and \cite{soloveychik2014tyler}.
In particular, the authors of \cite{soloveychik2013group} focused
on the group symmetry structure and proved that it is geodesically
convex. As the Tyler's estimator can be defined alternatively as the
minimizer of a cost function that is also geodesically convex, it
is concluded that any local minimum of the cost function on a group
symmetry constraint set is a global minimum. A numerical algorithm
was also proposed to solve the constrained minimization problem. In
\cite{soloveychik2014tyler}, a convex structural constraint set was
studied and a generalized method of moments type covariance estimator,
COCA, was proposed. A numerical algorithm was also provided based
on semidefinite relaxation. It was proved that COCA is an asymptotically
consistent estimator. However, the algorithm suffers from the drawback
that the computational cost increases as either $N$ or $K$ grows.

In this paper, we formulate the structured covariance estimation problem
as the minimization of Tyler's cost function under the structural
constraint. Our work generalizes \cite{soloveychik2013group} by considering
a much larger family of structures, which includes the group symmetry
structure. Instead of attempting to obtain a global optimal solution,
which is a challenging task due to the non-convexity of the objective
function, we focus on devising algorithms that converge to a stationary
point of the problem. We first work out an algorithm framework for
the general convex structural constraint based on the majorization
minimization (MM) framework, where a sequence of convex programming
is required to be solved. Then we consider several special cases that
appear frequently in practical applications. By exploiting specific
problem structures, the algorithm is particularized, significantly
reducing the computational load. We further discuss in the end two
types of widely studied non-convex structures that turn out to be
computationally tractable under the MM framework; one of them being
the Kronecker structure and the other one being the spiked covariance
structure. Although theoretically the algorithms can only be proved
to converge to a stationary point, for all the specific structures
that are considered in this paper, it is observed that the proposed
algorithms converge to a unique point (in $\mathbf{R}$) with random
initialization in the numerical simulations.

The paper is organized as follows. In Section II, we introduce the
robust covariance estimation problem with structural constraint and
derive a majorization minimization based algorithm framework for the
general convex structure. Several special cases are considered in
Section III, where the algorithm is particularized obtaining higher
efficiency by considering the specific form of the structure. Section
IV discusses the Kronecker structure and the spiked covariance structure,
which are non-convex but algorithmically tractable. Numerical results
are presented in Section V and we conclude in Section VI.

\section{Tyler's Estimator with Structural Constraint}

Consider a number of $N$ samples $\left\{ {\bf x}_{1},\ldots,{\bf x}_{N}\right\} $
in $\mathbb{R}^{K}$ drawn independently from an elliptical underlying
distribution with density function as follows:
\begin{equation}
f\left({\bf x}\right)=\det\left({\bf R}_{0}\right)^{-\frac{1}{2}}g\left({\bf x}^{T}{\bf R}_{0}^{-1}{\bf x}\right),\label{eq:-1}
\end{equation}
where ${\bf R}_{0}\in\mathbb{S}_{++}^{K}$ is the scatter parameter
that is proportional to the covariance matrix if it exists, and $g\left(\cdot\right)$
characterizes the shape of the distribution. Tyler's estimator for
${\bf R}_{0}$ is defined as the solution to the following fixed-point
equation:
\begin{equation}
{\bf R}=\frac{K}{N}\sum_{i=1}^{N}\frac{{\bf x}_{i}{\bf x}_{i}^{T}}{{\bf x}_{i}^{T}{\bf R}^{-1}{\bf x}_{i}},\label{eq:Tyler-estimator}
\end{equation}
which can be interpreted as a weighted sum of rank one matrices ${\bf x}_{i}{\bf x}_{i}^{T}$
with the weight decreasing as ${\bf x}_{i}$ gets farther from the
center. It is known that if ${\bf x}$ is elliptically distributed,
then the normalized random variable ${\bf s}=\frac{{\bf x}}{\left\Vert {\bf x}\right\Vert _{2}}$
follows an angular central Gaussian distribution with the probability
density function (pdf) taking the form
\begin{equation}
f\left({\bf s}\right)\propto\det\left({\bf R}_{0}\right)^{-\frac{1}{2}}\left({\bf s}^{T}{\bf R}_{0}^{-1}{\bf s}\right)^{-\frac{K}{2}}.\label{eq:ACG-pdf}
\end{equation}
Tyler's estimator coincides with the maximum likelihood estimator
(MLE) of ${\bf R}_{0}$ by fitting the normalized samples $\left\{ {\bf s}_{i}\right\} $
to $f\left({\bf s}\right)$. In other words, the estimator $\hat{{\bf R}}$
is the minimizer of the following cost function
\begin{equation}
L\left({\bf R}\right)=\log\det\left({\bf R}\right)+\frac{K}{N}\sum_{i=1}^{N}\log\left({\bf x}_{i}^{T}{\bf R}^{-1}{\bf x}_{i}\right)\label{eq:Tyler-cost-function}
\end{equation}
on the positive definite cone $\mathbb{S}_{++}^{K}$. The estimator
is proved to be consistent and asymptotically normal with the variance
independent of $g\left(\cdot\right)$. Furthermore, it is a minimax
robust covariance estimator with the underlying distribution being
elliptically symmetric \cite{Tyler1987}.

It has been noticed that in some applications, the covariance matrix
possesses a certain structure and taking account this information
into the estimation yields a better estimate of ${\bf R}_{0}$ \cite{Parikshit2011group,werner2008estimation,williams1993robust,Wirfalt2014kronecker}.
Motivated by this idea, we focus on the problem of including a prior
structure information into the Tyler's estimator to improve its estimation
accuracy. To formulate the problem, we assume that ${\bf R}_{0}$
is constrained in a non-empty set $\mathcal{S}$ that is the intersection
of a closed set, which characterizes the covariance structure, and
the positive semidefinite cone $\mathbb{S}_{+}^{K}$, and then proceed
to solve the optimization problem:
\begin{equation}
\begin{aligned} & \underset{{\bf R}}{\text{minimize}} &  & \log\det\left({\bf R}\right)+\frac{K}{N}\sum_{i=1}^{N}\log\left({\bf x}_{i}^{T}{\bf R}^{-1}{\bf x}_{i}\right)\\
 & \text{\text{subject to}} &  & {\bf R}\in\mathcal{S}.
\end{aligned}
\label{eq:P-ConstrainedTyler}
\end{equation}
The minimizer $\hat{{\bf R}}$ of the above problem is the one in
the structural set $\mathcal{S}$ that maximizes the likelihood of
the normalized samples $\left\{ {\bf s}_{i}\right\} $.

Throughout the paper, we make the following assumption.\textbf{}\\
\textbf{Assumption 1}: The cost function $L\left({\bf R}_{t}\right)\to+\infty$
when the sequence $\left\{ {\bf R}_{t}\right\} $ tends to a singular
limit point of the constraint set $\mathcal{S}$.

Under this assumption, the case that ${\bf R}$ is singular can be
excluded in the analysis of the algorithms hereafter.

Note that the assumption\textbf{}\\
\textbf{Assumption 2}: $f\left({\bf x}\right)$ is a continuous probability
distribution, and $N>K$ ,

implies $L\left({\bf R}_{t}\right)\to+\infty$ whenever ${\bf R}_{t}$
tends to the boundary of the positive semidefinite cone $\mathbb{S}_{+}^{K}$
with probability one \cite{Kent1988} . It is therefore also a sufficient
condition for the assumption to be held as $\mathcal{S}\subseteq\mathbb{S}_{+}^{K}$.

Problem (\ref{eq:P-ConstrainedTyler}) is difficult to solve for two
reasons. First, the constraint set $\mathcal{S}$ is too general to
tackle. Second, even if $\mathcal{S}$ possesses a nice property such
as convexity, the objective function is still non-convex. Instead
of trying to find the global minimizer, which appears to be too ambitious
for the reasons pointed out above, we aim at devising efficient algorithms
that are capable of finding a stationary point of (\ref{eq:P-ConstrainedTyler}).
We rely on the MM framework to derive the algorithms, which is briefly
stated next for completeness.

\subsection{The Majorization Minimization Algorithm}

For a general optimization problem
\begin{equation}
\begin{aligned} & \underset{{\bf x}}{\text{minimize}} &  & h\left({\bf x}\right)\\
 & \text{subject to} &  & {\bf x}\in\mathcal{X},
\end{aligned}
\label{eq:general-opt-p}
\end{equation}
where $\mathcal{X}$ is a closed convex set, the MM algorithm finds
a stationary point of (\ref{eq:general-opt-p}) by successively solving
a sequence of simpler optimization problems. The iterative algorithm
starts at some arbitrary feasible initial point ${\bf x}_{0}$, and
at the $\left(t+1\right)$-th iteration the update of ${\bf x}$ is
given by
\[
{\bf x}_{t+1}=\arg\min_{{\bf x}\in\mathcal{X}}g\left({\bf x}|{\bf x}_{t}\right),
\]
with the surrogate function $g\left({\bf x}|{\bf x}_{t}\right)$ satisfying
the following assumptions:
\begin{eqnarray}
h\left({\bf x}_{t}\right) & = & g\left({\bf x}_{t}|{\bf x}_{t}\right),\ \forall{\bf x}_{t}\in\mathcal{X}\nonumber \\
h\left({\bf x}\right) & \leq & g\left({\bf x}|{\bf x}_{t}\right),\ \forall{\bf x},{\bf x}_{t}\in\mathcal{X}\label{eq:MM-procedure}\\
h'\left({\bf x}_{t};{\bf d}\right) & = & g'\left({\bf x}_{t};{\bf d}|{\bf x}_{t}\right),\ \forall{\bf x}_{t}+{\bf d}\in\mathcal{X},\nonumber
\end{eqnarray}
where $h'\left({\bf x};{\bf d}\right)$ stands for the directional
derivative of $h\left(\cdot\right)$ at ${\bf x}$ along the direction
${\bf d}$, and $g\left({\bf x}|{\bf x}_{t}\right)$ is continuous
in both ${\bf x}$ and ${\bf x}_{t}$.

It is proved in \cite{razaviyayn2013unified} that any limit point
of the sequence $\left\{ {\bf x}_{t}\right\} $ generated by the MM
algorithm is a stationary point of problem (\ref{eq:general-opt-p}).
If it is further assumed that the initial level set $\left\{ {\bf x}|h\left({\bf x}\right)\leq h\left({\bf x}_{0}\right)\right\} $
is compact, then a stronger statement, as follows, can be made:
\[
\lim_{t\to+\infty}d\left({\bf x}_{t},\mathcal{X}^{\star}\right)=0,
\]
where $\mathcal{X}^{\star}$ stands for the set of all stationary
points of (\ref{eq:general-opt-p}).

The idea of majorizing $h\left({\bf x}\right)$ by a surrogate function
can also be applied blockwise. Specifically, ${\bf x}$ is partitioned
into $m$ blocks as ${\bf x}=\left({\bf x}^{\left(1\right)},\ldots,{\bf x}^{\left(m\right)}\right)$,
where each $n_{i}$-dimensional block ${\bf x}^{\left(i\right)}\in\mathcal{X}_{i}$
and $\mathcal{X}=\prod_{i=1}^{m}\mathcal{X}_{i}$.

At the $\left(t+1\right)$-th  iteration, ${\bf x}^{\left(i\right)}$
is updated by solving the following problem:
\begin{equation}
\begin{aligned} & \underset{{\bf x}^{\left(i\right)}}{\text{minimize}} &  & g_{i}\left({\bf x}^{\left(i\right)}|{\bf x}_{t}\right)\\
 & \text{subject to} &  & {\bf x}^{\left(i\right)}\in\mathcal{X}_{i}
\end{aligned}
\label{eq:MM-subprob}
\end{equation}
with $i=\left(t\ {\rm mod}\ m\right)+1$ and the continuous surrogate
function $g_{i}\left({\bf x}^{\left(i\right)}|{\bf x}_{t}\right)$
satisfying the following properties:
\begin{eqnarray*}
h\left({\bf x}_{t}\right) & = & g_{i}\left({\bf x}_{t}^{\left(i\right)}|{\bf x}_{t}\right),\\
h\left({\bf x}_{t}^{\left(1\right)},\ldots,{\bf x}^{\left(i\right)},\ldots,{\bf x}_{t}^{\left(m\right)}\right) & \leq & g_{i}\left({\bf x}^{\left(i\right)}|{\bf x}_{t}\right)\ \forall{\bf x}^{\left(i\right)}\in\mathcal{X}_{i},\\
h'\left({\bf x}_{t};{\bf d}_{i}^{0}\right) & = & g_{i}'\left({\bf x}_{t}^{\left(i\right)};{\bf d}_{i}|{\bf x}_{t}\right)\\
 &  & \forall{\bf x}_{t}^{\left(i\right)}+{\bf d}_{i}\in\mathcal{X}_{i},\\
 &  & \ {\bf d}_{i}^{0}\triangleq\left({\bf 0};\ldots;{\bf d}_{i};\ldots;{\bf 0}\right).
\end{eqnarray*}
In short, at each iteration, the block MM applies the ordinary MM
algorithm to one block while keeping the value of the other blocks
fixed. The blocks are updated in cyclic order.

In the rest of this paper, we are going to derive the specific form
of the surrogate function $g\left({\bf R}|{\bf R}_{t}\right)$ based
on a detailed characterization of various kinds of $\mathcal{S}$.
In addition, we are going to show how the algorithm can be particularized
at a lower computational cost with a finer structure of $\mathcal{S}$
available. Before moving to the algorithmic part, we first compare
our formulation with several related works in the literature.

\subsection{Related Works}

In \cite{williams1993robust}, the authors derived a minimax robust
covariance estimator assuming that $f\left({\bf x}\right)$ is a corrupted
Gaussian distribution with noise that belongs to the $\varepsilon$-contamination
class and the Kolmogorov class. The estimator is defined as the solution
of a constrained optimization problem similar to (\ref{eq:P-ConstrainedTyler}),
but with a different cost function. Apart from the distinction that
the family of distributions we consider is the set of elliptical distributions,
the focus of our work, which completely differs from \cite{williams1993robust},
is on developing efficient numerical algorithms for different types
of structural constraint set $\mathcal{S}$.

Two other closely related works are \cite{soloveychik2013group} and
\cite{soloveychik2014tyler}. In \cite{soloveychik2013group}, the
authors have investigated a special case of (\ref{eq:P-ConstrainedTyler}),
where $\mathcal{S}$ is the set of all positive semidefinite matrices
with group symmetry structure. It has been shown that both $L\left({\bf R}\right)$
and the group symmetry constraint are geodesically convex, therefore
any local minimizer of (\ref{eq:P-ConstrainedTyler}) is global. Several
examples, including the circulant and persymmetry structure, have
been proven to be a special case of the group symmetry constraint.
A numerical algorithm has also been provided that decreases the cost
function monotonically. Our work includes the group symmetry structure
as a special case since the constraint is linear, and provides an
alternative algorithm to solve the problem.

In \cite{soloveychik2014tyler}, the authors have considered imposing
convex constraint on Tyler's estimator. A generalized method of moment
type estimator based on semidefinite relaxation defined as the solution
of the following problem:
\begin{equation}
\begin{aligned} & \underset{{\bf R}\in\mathcal{S},d_{i}}{\text{minimize}} &  & \left\Vert {\bf R}-\frac{1}{N}\sum_{i=1}^{N}d_{i}{\bf x}_{i}{\bf x}_{i}^{T}\right\Vert \\
 & \text{subject to} &  & {\bf R}\succeq\frac{1}{K}d_{i}{\bf x}_{i}{\bf x}_{i}^{T},\ \forall i=1,\ldots,N,\\
 &  &  & d_{i}>0,\ \forall i=1,\ldots,N,
\end{aligned}
\label{eq:COCA}
\end{equation}
was proposed and proved to be asymptotically consistent. Nevertheless,
the number of constraints grows linearly in $N$ and as it was pointed
out in the paper, the algorithm becomes computationally demanding
either when the problem dimension $K$ or the number of samples $N$
is large. On the contrary, our algorithm based on formulation (\ref{eq:P-ConstrainedTyler})
is less affected by the number of samples $N$ and is therefore more
computationally tractable.

\section{Tyler's Estimator with Convex structural constraint}

In this section, we are going to derive a general algorithm for problem
(\ref{eq:P-ConstrainedTyler}) with $\mathcal{S}$ being a closed
convex subset of $\mathbb{S}_{+}^{K}$, which enjoys a wide range
of applications. For instance, the Toeplitz structure can be imposed
on the covariance matrix of the received signal in direction-of-arrival
estimation (DOA) problems. Banding is also considered as a way of
regularizing a covariance matrix whose entries decay fast as they
get far away from the main diagonal.

Since $\mathcal{S}$ is closed and convex, constructing a convex surrogate
function $g\left({\bf R}|{\bf R}_{t}\right)$ for $L\left({\bf R}\right)$
turns out to be a natural idea since then ${\bf R}_{t+1}$ can be
found via
\begin{equation}
{\bf R}_{t+1}=\arg\min_{{\bf R}\in\mathcal{S}}\ g\left({\bf R}|{\bf R}_{t}\right),\label{eq:alg-general cvx set}
\end{equation}
which is a convex programming.
\begin{prop}
\label{prop:full-mm-bound}At any ${\bf R}_{t}\succ{\bf 0}$, the
objective function $L\left({\bf R}\right)$ can be upperbounded by
the convex surrogate function
\begin{equation}
g\left({\bf R}|{\bf R}_{t}\right)=\text{Tr}\left({\bf R}_{t}^{-1}{\bf R}\right)+\frac{K}{N}\sum_{i=1}^{N}\frac{{\bf x}_{i}^{T}{\bf R}^{-1}{\bf x}_{i}}{{\bf x}_{i}^{T}{\bf R}_{t}^{-1}{\bf x}_{i}}+\text{const.}\label{eq:full-mm}
\end{equation}
with equality achieved at ${\bf R}_{t}$.\end{prop}
\begin{IEEEproof}
Since $\log\det\left(\cdot\right)$ is concave, $\log\det\left({\bf R}\right)$
can be upperbounded by its first order Taylor expansion at ${\bf R}_{t}$:
\begin{equation}
\log\det\left({\bf R}\right)\leq\log\det\left({\bf R}_{t}\right)+{\rm Tr}\left({\bf R}_{t}^{-1}{\bf R}\right)-K\label{eq:log-det-upperbound}
\end{equation}
with equality achieved at ${\bf R}_{t}$.

Also, by the concavity of the $\log\left(\cdot\right)$ function we
have
\begin{equation}
\log\left(x\right)\leq\log a+\frac{x}{a}-1,\ \forall a>0,\label{eq:log-upperbound}
\end{equation}
which leads to the bound
\[
\log\left({\bf x}_{i}^{T}{\bf R}^{-1}{\bf x}_{i}\right)\leq\frac{{\bf x}_{i}^{T}{\bf R}^{-1}{\bf x}_{i}}{{\bf x}_{i}^{T}{\bf R}_{t}^{-1}{\bf x}_{i}}+\log\left({\bf x}_{i}^{T}{\bf R}_{t}^{-1}{\bf x}_{i}\right)-1
\]
with equality achieved at ${\bf R}_{t}$.
\end{IEEEproof}
The variable ${\bf R}$ then can be updated as (\ref{eq:alg-general cvx set})
with surrogate function (\ref{eq:full-mm}).

By the convergence result of the MM algorithm, it can be concluded
that every limit point of the sequence $\left\{ {\bf R}_{t}\right\} $
is a stationary point of problem (\ref{eq:P-ConstrainedTyler}). Note
that for all of the structural constraints that we are going to consider
in this work, the set $\mathcal{S}$ possesses the property that
\begin{equation}
{\bf R}\in\mathcal{S}\ \text{iff}\ r{\bf R}\in\mathcal{S},\ \forall r>0.\label{eq:set-scale-invariancy}
\end{equation}
Since the cost function $L\left({\bf R}\right)$ is scale-invariant
in the sense that $L\left({\bf R}\right)=L\left(r{\bf R}\right)$,
we can add a trace normalization step after the update of ${\bf R}_{t}$
without affecting the value of the objective function. The algorithm
for a general convex structural set is summarized in Algorithm \ref{SCA}.

\begin{algorithm}
\caption{\label{SCA}Robust covariance estimation under convex structure}
\begin{algorithmic}[1]
\State Set $t=0$, initialize ${\bf R}_t$ to be any positive definite matrix.
\Repeat
\State Compute ${\bf M}_{t}=\frac{K}{N}\sum_{i=1}^{N}\frac{{\bf x}_{i}{\bf x}_{i}^{T}}{{\bf x}_{i}^{T}{\bf R}_{t}^{-1}{\bf x}_{i}}$.
\State Update ${\bf R}_{t+1}$ as \begin{eqnarray}\tilde{{\bf R}}_{t+1}&=&\arg\min_{{\bf R}\in\mathcal{S}}{\rm Tr}\left({\bf R}_{t}^{-1}{\bf R}\right)+\textrm{Tr}\left({\bf M}_{t}{\bf R}^{-1}\right)\label{eq:update R}\\{\bf R}_{t+1}&=&\tilde{{\bf R}}_{t+1}/\text{Tr}\left(\tilde{{\bf R}}_{t+1}\right).\end{eqnarray}
\State $t \gets t+1$.
\Until{Some convergence criterion is met}
\end{algorithmic}
\end{algorithm}
\begin{prop}
If the set $\mathcal{S}$ satisfies (\ref{eq:set-scale-invariancy}),
then the sequence $\left\{ {\bf R}_{t}\right\} $ generated by Algorithm
\ref{SCA} satisfies
\begin{equation}
\lim_{t\to\infty}d\left({\bf R}_{t},\mathcal{S}^{\star}\right)=0,\label{eq:converge to stationary set}
\end{equation}
where $\mathcal{S}^{\star}$ is the set of stationary points of problem
(\ref{eq:P-ConstrainedTyler}).\end{prop}
\begin{IEEEproof}
Since the objective function $L\left({\bf R}\right)$ is scale-invariant,
and the constraint set satisfies (\ref{eq:set-scale-invariancy}),
solving (\ref{eq:P-ConstrainedTyler}) is equivalent to solving
\[
\begin{aligned} & \underset{{\bf R}\in\mathcal{S}}{\text{minimize}} &  & \log\det\left({\bf R}\right)+\frac{K}{N}\sum_{i=1}^{N}\log\left({\bf x}_{i}^{T}{\bf R}^{-1}{\bf x}_{i}\right)\\
 & \text{\text{subject to}} &  & {\rm Tr}\left({\bf R}\right)=1.
\end{aligned}
\]
The conclusion follows by a similar argument to Proposition 17 in
\cite{sun2014RegularTyler}.
\end{IEEEproof}

\subsection{General Linear Structure }

In this subsection we further assume that the set $\mathcal{S}$ is
the intersection of $\mathbb{S}_{+}^{K}$ and an affine set $\mathcal{A}$.
The following lemma shows that in this case, the update of ${\bf R}$
(eqn. \eqref{eq:update R}) can be recast as an SDP.
\begin{lem}
\label{lem:equivalent SDP} Problem \eqref{eq:update R} is equivalent
to
\begin{equation}
\begin{aligned} & \underset{{\bf S},{\bf R}\in\mathcal{S}}{\textrm{minimize}} &  & \textrm{Tr}\left({\bf R}_{t}^{-1}{\bf R}\right)+\textrm{Tr}\left({\bf M}_{t}{\bf S}\right)\\
 & \textrm{subject to} &  & \left[\begin{array}{cc}
{\bf S} & {\bf I}\\
{\bf I} & {\bf R}
\end{array}\right]\succeq{\bf 0},
\end{aligned}
\label{eq:equivalent SDP}
\end{equation}
in the sense that if $\left({\bf S}^{\star},{\bf R}^{\star}\right)$
solves (\ref{eq:equivalent SDP}), then ${\bf R}^{\star}$ solves
\eqref{eq:update R}.\end{lem}
\begin{IEEEproof}
Problem \eqref{eq:update R} can be written equivalently as
\[
\begin{aligned} & \underset{{\bf S},{\bf R}\in\mathcal{S}}{\textrm{minimize}} &  & \textrm{Tr}\left({\bf R}_{t}^{-1}{\bf R}\right)+\textrm{Tr}\left({\bf M}_{t}{\bf S}\right)\\
 & \textrm{subject to} &  & {\bf S}={\bf R}^{-1}.
\end{aligned}
\]
Now we relax the constraint ${\bf S}={\bf R}^{-1}$ as ${\bf S}\succeq{\bf R}^{-1}$.
By the Schur complement lemma for a positive semidefinite matrix,
if ${\bf R}\succ{\bf 0}$, then ${\bf S}\succeq{\bf R}^{-1}$ is equivalent
to
\[
\left[\begin{array}{cc}
{\bf S} & {\bf I}\\
{\bf I} & {\bf R}
\end{array}\right]\succeq{\bf 0}.
\]
Therefore (\ref{eq:equivalent SDP}) is a convex relaxation of \eqref{eq:update R}.

The relaxation is tight since ${\rm Tr}\left({\bf M}_{t}{\bf S}\right)\geq{\rm Tr}\left({\bf M}_{t}{\bf R}^{-1}\right)$
if ${\bf M}_{t}\succeq{\bf 0}$ and ${\bf S}\succeq{\bf R}^{-1}$.
\end{IEEEproof}
Lemma \ref{lem:equivalent SDP} reveals that for linear structural
constraint, Algorithm \ref{SCA}  can be particularized as solving
a sequence of SDPs.

An application is the case that ${\bf R}$ can be parametrized as
\begin{equation}
{\bf R}=\sum_{j=1}^{L}a_{j}{\bf B}_{j}\label{eq:linear structure}
\end{equation}
with $a_{j}\in\mathbb{R}$ being the variable and ${\bf B}_{j}\in\mathbb{R}^{K\times K}$
being the corresponding given basis matrix, and ${\bf R}$ is constrained
to be in $\mathbb{S}_{+}^{K}$. Using expression (\ref{eq:linear structure}),
the minimization problem \eqref{eq:equivalent SDP} can be simplified
as
\begin{equation}
\begin{aligned} & \underset{{\bf S},\left\{ a_{j}\right\} }{\textrm{minimize}} &  & \sum_{j=1}^{L}a_{j}\textrm{Tr}\left({\bf R}_{t}^{-1}{\bf B}_{j}\right)+\textrm{Tr}\left({\bf M}_{t}{\bf S}\right)\\
 & \textrm{subject to} &  & \left[\begin{array}{cc}
{\bf S} & {\bf I}\\
{\bf I} & \sum_{j=1}^{L}a_{j}{\bf B}_{j}
\end{array}\right]\succeq{\bf 0}.
\end{aligned}
\label{eq:LMI}
\end{equation}

\section{Tyler's Estimator with Special Convex Structures}

Having introduced the general algorithm framework for a convex structure
in the previous section, we are going to discuss in detail some convex
structures that arise frequently in signal processing related fields,
and show that by exploiting the problem structure the algorithm can
be particularized with a significant reduction in the computational
load.

\subsection{Sum of Rank-One Matrices Structure}

The structure set $\mathcal{S}$ that we study in this part is

\begin{equation}
\mathcal{S}=\left\{ {\bf R}|{\bf R}=\sum_{j=1}^{L}p_{j}{\bf a}_{j}{\bf a}_{j}^{H},\ p_{j}\geq0\right\} ,\label{eq:decomposable structure}
\end{equation}
where the ${\bf a}_{j}$'s are known vectors in $\mathbb{C}^{K}$.
The matrix ${\bf R}$ can be interpreted as a weighted sum of given
matrices ${\bf a}_{j}{\bf a}_{j}^{H}$.

As an example application where structure (\ref{eq:decomposable structure})
appears, consider the following signal model
\begin{equation}
{\bf x}={\bf A}\boldsymbol{\beta}+\boldsymbol{\varepsilon},\label{eq:signal model}
\end{equation}
where ${\bf A}=\left[{\bf a}_{1},\ldots,{\bf a}_{L}\right]$. Assuming
that the signal $\boldsymbol{\beta}$ and noise $\boldsymbol{\varepsilon}$
are zero-mean random variables and any two elements of them are uncorrelated,
then the covariance matrix of ${\bf x}$ takes the form
\begin{equation}
{\rm Cov}\left({\bf x}\right)=\sum_{j=1}^{L}p_{j}{\bf a}_{j}{\bf a}_{j}^{H}+\boldsymbol{\Sigma},\label{eq:signal-noise model}
\end{equation}
where $p_{j}={\rm Var}\left(\beta_{j}\right)$ is the signal variance
and $\boldsymbol{\Sigma}={\rm diag}\left(\sigma_{1},\ldots,\sigma_{K}\right)$
is the noise covariance matrix.

Define ${\bf p}=\left[p_{1},\ldots p_{L}\right]^{T}$ and ${\bf P}={\rm diag}\left({\bf p}\right)$,
then ${\bf R}$ can be written compactly as ${\bf R}={\bf A}{\bf P}{\bf A}^{H}+\boldsymbol{\Sigma}$.
Further define
\begin{equation}
\begin{aligned}\tilde{{\bf P}} & ={\rm diag}\left(p_{1},\ldots,p_{L},\sigma_{1},\ldots,\sigma_{K}\right)\\
\tilde{{\bf A}} & =\left[{\bf A},{\bf I}\right]
\end{aligned}
\label{eq:augment basis}
\end{equation}
 then ${\bf R}=\tilde{{\bf A}}\tilde{{\bf P}}\tilde{{\bf A}}^{H}$.
Therefore, without loss of generality, we can focus on the expression
${\bf R}={\bf A}{\bf P}{\bf A}^{H}$, assuming that every $K$ columns
of ${\bf A}$ are linearly independent and $L>K$.

Note that in example (\ref{eq:signal model}), ${\bf R}$ is complex-valued
and problem (\ref{eq:P-ConstrainedTyler}), which is formulated based
on the real-valued elliptical distribution $f\left({\bf x}\right)$,
needs to be modified to
\begin{equation}
\begin{aligned} & \underset{{\bf R},{\bf P}\succeq{\bf 0}}{\text{minimize}} &  & \log\det\left({\bf R}\right)+\frac{K}{N}\sum_{i=1}^{N}\log\left({\bf x}_{i}^{H}{\bf R}^{-1}{\bf x}_{i}\right)\\
 & \text{subject to} &  & {\bf R}={\bf A}{\bf P}{\bf A}^{H},
\end{aligned}
\label{eq:P-sum-of-rank-one}
\end{equation}
where the ${\bf x}_{i}$'s are assumed follow a complex-valued elliptical
distribution instead.

Since ${\bf R}$ is linear in the $p_{j}$'s, Algorithm \ref{SCA}
can be applied. In the following, we are going to provide a more efficient
algorithm by substituting ${\bf R}={\bf A}{\bf P}{\bf A}^{H}$ into
the objective function $L\left({\bf R}\right)$ and applying the MM
procedure with ${\bf P}$ being the variable.
\begin{prop}
\label{lem:bound-P}At any ${\bf P}_{t}\succ{\bf 0}$, the objective
function
\begin{equation}
L\left({\bf P}\right)=\log\det\left({\bf A}{\bf P}{\bf A}^{H}\right)+\frac{K}{N}\sum_{i=1}^{N}\log\left({\bf x}_{i}^{H}\left({\bf A}{\bf P}{\bf A}^{H}\right)^{-1}{\bf x}_{i}\right)\label{eq:L(p)}
\end{equation}
can be upperbounded by the surrogate function
\begin{equation}
g\left({\bf P}|{\bf P}_{t}\right)={\bf w}_{t}^{T}{\bf p}+{\bf d}_{t}^{T}{\bf p}^{-1}+{\rm const}.\label{eq:surrogate function P}
\end{equation}
with equality achieved at ${\bf P}={\bf P}_{t}$, where ${\bf p}^{-1}$
stands for the element-wise inverse of ${\bf p}$, and
\begin{equation}
\begin{aligned}{\bf R}_{t} & ={\bf A}{\bf P}_{t}{\bf A}^{H}\\
{\bf M}_{t} & =\frac{K}{N}\sum_{i=1}^{N}\frac{{\bf x}_{i}{\bf x}_{i}^{T}}{{\bf x}_{i}^{T}{\bf R}_{t}^{-1}{\bf x}_{i}}\\
{\bf w}_{t} & ={\rm diag}\left({\bf A}^{H}{\bf R}_{t}^{-1}{\bf A}\right)\\
{\bf d}_{t} & ={\rm diag}\left({\bf P}_{t}{\bf A}^{H}{\bf R}_{t}^{-1}{\bf M}_{t}{\bf R}_{t}^{-1}{\bf A}{\bf P}_{t}\right).
\end{aligned}
\label{eq:parameters}
\end{equation}
\end{prop}
\begin{IEEEproof}
First, observe that inequalities (\ref{eq:log-det-upperbound}) and
(\ref{eq:log-upperbound}) imply that
\begin{eqnarray}
L\left({\bf P}\right) & \leq & {\bf w}_{t}^{T}{\bf p}+{\rm Tr}\left({\bf M}_{t}{\bf R}^{-1}\right)+{\rm const.}\label{eq:-2}
\end{eqnarray}
with equality achieved at ${\bf P}={\bf P}_{t}$.

Assume that ${\bf P}\succ{\bf 0}$, from the identity
\begin{align*}
{\bf S}= & \left[\begin{array}{cc}
{\bf R}_{t}^{-1}{\bf A}{\bf P}_{t}{\bf P}^{-1}{\bf P}_{t}{\bf A}^{H}{\bf R}_{t}^{-1} & {\bf I}\\
{\bf I} & {\bf A}{\bf P}{\bf A}^{H}
\end{array}\right]\\
= & \left[\begin{array}{c}
{\bf R}_{t}^{-1}{\bf A}{\bf P}_{t}{\bf P}^{-1/2}\\
{\bf A}{\bf P}^{1/2}
\end{array}\right]\left[\begin{array}{cc}
{\bf P}^{-1/2}{\bf P}_{t}{\bf A}^{H}{\bf R}_{t}^{-1} & {\bf P}^{1/2}{\bf A}^{H}\end{array}\right],
\end{align*}
we know that ${\bf S}\succeq{\bf 0}$. By the Schur complement, ${\bf S}\succeq{\bf 0}$
is equivalent to
\begin{equation}
{\bf R}_{t}^{-1}{\bf A}{\bf P}_{t}{\bf P}^{-1}{\bf P}_{t}{\bf A}^{H}{\bf R}_{t}^{-1}\succeq\left({\bf A}{\bf P}{\bf A}^{H}\right)^{-1}.\label{eq:matrix-inequality}
\end{equation}
Since ${\bf M}_{t}\succeq{\bf 0}$, we have
\begin{equation}
{\rm Tr}\left({\bf M}_{t}{\bf R}^{-1}\right)\leq{\rm Tr}\left({\bf M}_{t}{\bf R}_{t}^{-1}{\bf A}{\bf P}_{t}{\bf P}^{-1}{\bf P}_{t}{\bf A}^{H}{\bf R}_{t}^{-1}\right)\label{eq:upperbound-trace-inverse}
\end{equation}
with equality achieved at ${\bf P}={\bf P}_{t}$.

Since ${\bf R}\succ{\bf 0}$, the left hand side of (\ref{eq:upperbound-trace-inverse})
is finite. Therefore (\ref{eq:upperbound-trace-inverse}) is also
valid for ${\bf P}\succeq{\bf 0}$. Substituting (\ref{eq:upperbound-trace-inverse})
into (\ref{eq:-2}) yields the surrogate function (\ref{eq:surrogate function P}).
\end{IEEEproof}
The update of ${\bf P}$ then can be found in closed-form as

\begin{equation}
\begin{aligned}\left(p_{j}\right)_{t+1} & =\sqrt{\left(d_{j}\right)_{t}/\left(w_{j}\right)_{t}}.\end{aligned}
\label{eq:fixed-point-itr-p}
\end{equation}
The algorithm is summarized in Algorithm \ref{Acc-LIKES}.

Compared to Algorithm \ref{SCA}, in which the minimization problem
(\ref{eq:alg-general cvx set}) has no closed-form solution and typically
requires an iterative algorithm, the new algorithm only requires a
single loop iteration in ${\bf p}$ and is expected to converge faster.

\begin{algorithm}   \caption{\label{Acc-LIKES}Robust covariance estimation under sum of rank-one matrices structure}     \begin{algorithmic}[1]       \State Set $t=0$, initialize ${\bf p}_t$ to be any positive vector.
\Repeat
\State $\tilde{{\bf R}}_t = {\bf A}{\bf P}_t{\bf A}^H$, ${\bf R}_t=\tilde{{\bf R}}_t/\text{Tr}\left(\tilde{{\bf R}}_t\right).$
\State Compute  ${\bf M}_t$, ${\bf w}_t$, ${\bf d}_t$  with \eqref{eq:parameters}       \State $\left(p_j\right)_{t+1}=\sqrt{\left(d_j\right)_{t}/\left(w_j\right)_t}$       \State $t \gets t+1$.       \Until{some convergence criterion is met}     \end{algorithmic}   \end{algorithm}

\subsection{Toeplitz Structure}

Consider the constraint set being the class of real-valued positive
semidefinite Toeplitz matrices $T_{K}$. If ${\bf R}\in T_{K}$, then
it can be completely determined by its first row%
\footnote{Following the convention, the indices for the Toeplitz structure start
from $0$.%
} $\left[r_{0},\ldots,r_{K-1}\right]$.

In this subsection, we are going to show that based on the technique
of circulant embedding, Algorithm \ref{Acc-LIKES} can be adopted
to solve the Toeplitz structure constrained problem at a lower cost
than applying the sequential SDP algorithm (Algorithm \ref{SCA}).

The idea of embedding a Toeplitz matrix as the upper-left part of
a larger circulant matrix has been discussed in \cite{fuhrmann1988existence,miller1987role,miller1986alternating}.
It was proved in \cite{dembo1989embedding} that any positive definite
Toeplitz matrix ${\bf R}$ of size $K\times K$ can be embedded in
a positive definite circulant matrix ${\bf C}$ of larger size $L\times L$
parametrized by its first row of the form
\[
\left[r_{0},r_{1},\ldots,r_{K-1},\ast,\ldots,\ast,r_{K-1},\ldots,r_{1}\right],
\]
where $\ast$ denotes some real number. ${\bf R}$ then can be written
as
\begin{equation}
{\bf R}=\left[\begin{array}{cc}
{\bf I}_{K} & {\bf 0}\end{array}\right]{\bf C}\left[\begin{array}{cc}
{\bf I}_{K} & {\bf 0}\end{array}\right]^{T}.\label{eq:-3}
\end{equation}
Clearly, for any fixed $L$, if ${\bf C}$ is positive semidefinite,
so is ${\bf R}$. However, the statement is false the other way around.
In other words, the set
\begin{equation}
T_{K}^{L}\triangleq\left\{ {\bf R}|{\bf R}=\left[\begin{array}{cc}
{\bf I}_{K} & {\bf 0}\end{array}\right]{\bf C}\left[\begin{array}{cc}
{\bf I}_{K} & {\bf 0}\end{array}\right]^{T},{\bf C}\in C_{L}\right\} ,\label{eq:approximate-toeplitz-set}
\end{equation}
where $C_{L}$ denotes the set of real-valued positive semidefinite
circulant matrices of size $L\times L$, is a subset of $T_{K}$.

Instead of $T_{K}$, we restrict the feasible set to be $T_{K}^{L}$
with $L\geq2K-1$. Since a symmetric circulant matrix can be diagonalized
by the Fourier matrix, if ${\bf R}\in T_{K}^{L}$ then it can be written
as
\begin{equation}
{\bf R}={\bf A}\text{diag}\left(p_{0},\ldots,p_{L-1}\right){\bf A}^{H},\label{eq:-4}
\end{equation}
where
\begin{equation}
{\bf A}=\left[\begin{array}{cc}
{\bf I}_{K} & {\bf 0}\end{array}\right]{\bf F}_{L},\label{eq:-5}
\end{equation}
with ${\bf F}_{L}$ being the normalized Fourier transform matrix
of size $L\times L$ and $p_{j}=p_{L-j},\ \forall j=1,\ldots,L-1$.

The robust covariance estimation problem over the restricted set of
Toeplitz matrices $T_{K}^{L}$ then takes the form
\begin{equation}
\begin{aligned} & \underset{{\bf R},{\bf P}\succeq{\bf 0}}{\text{minimize}} &  & \log\det\left({\bf R}\right)+\frac{K}{N}\sum_{i=1}^{N}\log\left({\bf x}_{i}^{H}{\bf R}^{-1}{\bf x}_{i}\right)\\
 & \text{subject to} &  & {\bf R}={\bf A}{\bf P}{\bf A}^{H}\\
 &  &  & p_{j}=p_{L-j},\ \forall j=1,\ldots,L-1,
\end{aligned}
\label{eq:P-approx-toeplitz-equivalent}
\end{equation}
which is the same as (\ref{eq:P-sum-of-rank-one}) except that the
last equality constraint on the $p_{j}$'s.

By Proposition \ref{lem:bound-P}, the inner minimization problem
takes the form
\begin{equation}
\begin{aligned} & \underset{{\bf p}\geq{\bf 0}}{\text{minimize}} &  & {\bf w}_{t}^{T}{\bf p}+{\bf d}_{t}^{T}{\bf p}^{-1}\\
 & \text{subject to} &  & p_{j}=p_{L-j},\ \forall j=1,\ldots,L-1.
\end{aligned}
\label{eq:P-inner-min-toeplitz}
\end{equation}
Note that by the property of the Fourier transform matrix, we have
${\bf a}_{j}=\bar{{\bf a}}_{L-j},\ \forall j=1,\ldots,L-1$, where
the upper bar stands for element-wise complex conjugate. As a result,
for $j=1,\ldots,L-1$,
\begin{equation}
\begin{aligned}\left(w_{j}\right)_{t} & =\left(w_{L-j}\right)_{t}\\
\left(d_{j}\right)_{t} & =\left(d_{L-j}\right)_{t},
\end{aligned}
\label{eq:coefficient-equal}
\end{equation}
which implies that the constraint $p_{j}=p_{L-j}$ will be satisfied
automatically.

The algorithm for the Toeplitz structure based on circulant embedding
is summarized in Algorithm \ref{Circulant-Embedding}. Notice that
Algorithm \ref{Circulant-Embedding} can be generalized easily to
noisy observations by the augmented representation (\ref{eq:augment basis}).

\begin{algorithm}   \caption{\label{Circulant-Embedding}Robust covariance estimation under  the Toeplitz structure (Circulant Embedding)}     \begin{algorithmic}[1]
\State Set $L$ to be an integer such that $L\geq 2K-1$.
\State Construct matrix ${\bf A}=\left[\begin{array}{cc}{\bf I}_{K} & {\bf 0}\end{array}\right]{\bf F}_{L}$
 \State Call Algorithm \ref{Acc-LIKES}.
   \end{algorithmic}   \end{algorithm}

\subsection{Banded Toeplitz Structure}

In addition to imposing the Toeplitz structure on the covariance matrix,
in some applications we can further require that the Toeplitz matrix
is $k$-banded, i.e., $r_{j}=0$ if $j>k$. For example, the covariance
matrix of a stationary moving average process of order $k$ satisfies
the above assumption. One may also consider banding the covariance
matrix if it is known in prior that the correlation of $x_{t}$ and
$x_{t-\tau}$ decreases as $\tau$ increases.

Based on the circulant embedding technique introduced in the last
subsection, the problem can be formulated as
\begin{equation}
\begin{aligned} & \underset{{\bf R},{\bf P}\succeq{\bf 0}}{\text{minimize}} &  & \log\det\left({\bf R}\right)+\frac{K}{N}\sum_{i=1}^{N}\log\left({\bf x}_{i}^{H}{\bf R}^{-1}{\bf x}_{i}\right)\\
 & \text{subject to} &  & {\bf R}={\bf A}{\bf P}{\bf A}^{H}\\
 &  &  & p_{j}=p_{L-j},\ \forall j=1,\ldots,L-1\\
 &  &  & r_{j}=0,\ \forall j=k+1,\ldots,K-1.
\end{aligned}
\label{eq:banded-toeplitz}
\end{equation}
By Proposition \ref{lem:bound-P}, the inner minimization problem
becomes
\begin{equation}
\begin{aligned} & \underset{{\bf p}\geq{\bf 0}}{\text{minimize}} &  & {\bf w}_{t}^{T}{\bf p}+{\bf d}_{t}^{T}{\bf p}^{-1}\\
 & \text{subject to} &  & p_{j}=p_{L-j},\ \forall j=1,\ldots,L-1\\
 &  &  & r_{j}=0,\ \forall j=k+1,\ldots,K-1,
\end{aligned}
\label{eq:P-inner-min-toeplitz-banded}
\end{equation}
which can be rewritten compactly as
\begin{equation}
\begin{aligned} & \underset{{\bf p}\geq{\bf 0}}{\text{minimize}} &  & {\bf w}_{t}^{T}{\bf p}+{\bf d}_{t}^{T}{\bf p}^{-1}\\
 & \text{subject to} &  & \left[\begin{array}{cc}
{\bf 0}_{\left(K-k-1\right)\times k+1} & {\bf I}_{K-k-1}\end{array}\right]{\bf A}{\bf p}={\bf 0}\\
 &  &  & p_{j}=p_{L-j},\ \forall j=1,\ldots,L-1.
\end{aligned}
\label{eq:P-dual-decompse}
\end{equation}
Define real-valued quantities
\begin{eqnarray}
\tilde{{\bf A}} & = & {\rm Re}\left\{ \left[\sqrt{2}{\bf a}_{0},{\bf a}_{1},\ldots,{\bf a}_{\left\lceil \frac{L-1}{2}\right\rceil }\right]\right\} \label{eq:A tilde}\\
\tilde{{\bf w}} & = & \left[\sqrt{2}w_{0},w_{1},\ldots,w_{\left\lceil \frac{L-1}{2}\right\rceil }\right]\label{eq:w tilde}\\
\tilde{{\bf d}} & = & \left[d_{0}/\sqrt{2},d_{1},\ldots,d_{\left\lceil \frac{L-1}{2}\right\rceil }\right],\label{eq:d tilde}
\end{eqnarray}
we have the equivalent problem
\begin{equation}
\begin{aligned} & \underset{\tilde{{\bf p}}\geq{\bf 0}}{\text{minimize}} &  & \tilde{{\bf w}}_{t}^{T}\tilde{{\bf p}}+\sum_{j=0}^{\left\lceil \frac{L-1}{2}\right\rceil }\tilde{d}_{j}/\tilde{p}_{j}\\
 & \text{subject to} &  & \tilde{{\bf A}}\tilde{{\bf p}}={\bf 0},
\end{aligned}
\label{eq:surrogate problem-baned-toeplitz}
\end{equation}
 where the variables $\tilde{{\bf p}}$ and ${\bf p}$ are related
by
\begin{equation}
\tilde{{\bf p}}=\left[p_{0}/\sqrt{2},p_{1},\ldots,p_{\left\lceil \frac{L-1}{2}\right\rceil }\right].\label{eq:p-ptilde}
\end{equation}
Compared to (\ref{eq:P-dual-decompse}), the equivalent problem has
a lower computational cost as both the number of variables and constraints
are reduced. The algorithm for the banded Toeplitz structure is summarized
in Algorithm \ref{banded-toeplitz}.

\begin{algorithm}   \caption{\label{banded-toeplitz}Robust covariance estimation under the Banded Toeplitz structure (Circulant Embedding)}     \begin{algorithmic}[1]
\State Set $L$ to be an integer such that $L\geq 2K-1$.
\State Construct matrix ${\bf A}=\left[\begin{array}{cc}{\bf I}_{K} & {\bf 0}\end{array}\right]{\bf F}_{L}$ and $\tilde {\bf A}$ with \eqref{eq:A tilde}.
 \State Set $t=0$, initialize ${\bf p}_t$ to be any positive vector.
  \Repeat       \State $\tilde {\bf R}_t={\bf A}{\bf P}_t{\bf A}^H$, ${\bf R}_{t}=\tilde{{\bf R}}_{t}/\text{Tr}\left(\tilde{{\bf R}}_{t}\right).$       \State Compute  ${\bf M}_t$, ${\bf w}_t$, ${\bf d}_t$  with \eqref{eq:parameters}.
\State Compute $\tilde {\bf w}$ and $\tilde {\bf d}$ with \eqref{eq:w tilde} and  \eqref{eq:d tilde}, and update $\tilde {\bf p}$ as the minimizer of
\eqref{eq:surrogate problem-baned-toeplitz}.
\State Compute $\bf p$ with \eqref{eq:p-ptilde}, ${\bf p}_t \gets {\bf p}$
\State $t \gets t+1$       \Until{some convergence criterion is met}     \end{algorithmic}   \end{algorithm}

\subsection{Convergence Analysis}

We consider Algorithm \ref{Acc-LIKES}, and the argument for Algorithms
\ref{Circulant-Embedding}, and \ref{banded-toeplitz} would be similar.

As Proposition \ref{lem:bound-P} requires ${\bf P}_{t}\succ{\bf 0}$,
we consider the following $\epsilon$-approximation of problem (\ref{eq:P-sum-of-rank-one}):
\begin{equation}
\begin{aligned} & \underset{{\bf R},{\bf p}\geq{\bf 0}}{\text{minimize}} &  & \log\det\left({\bf R}+\epsilon{\bf A}{\bf A}^{H}\right)\\
 &  &  & +\frac{K}{N}\sum_{i=1}^{N}\log\left({\bf x}_{i}^{H}\left({\bf R}+\epsilon{\bf A}{\bf A}^{H}\right)^{-1}{\bf x}_{i}\right)\\
 & \text{subject to} &  & {\bf R}={\bf A}{\bf P}{\bf A}^{H}
\end{aligned}
\label{eq:approx-sum-rank-one}
\end{equation}
with $\varepsilon>0$, where the upperbound derived in Proposition
\ref{lem:bound-P} can be applied for $\tilde{{\bf P}}\triangleq{\bf P}+\epsilon{\bf I}$.
Algorithm \ref{Acc-LIKES} can be easily modified to solve problem
(\ref{eq:approx-sum-rank-one}), and under Assumption 1, the limit
point of the sequence $\left\{ {\bf p}_{t}^{\epsilon}\right\} $ generated
by Algorithm \ref{Acc-LIKES} converges to the set of stationary points
of (\ref{eq:approx-sum-rank-one}).

That is, if $\left({\bf p}^{\epsilon}\right)^{\star}$ is a limit
point of $\left\{ {\bf p}_{t}^{\epsilon}\right\} $, then

\begin{equation}
\nabla L^{\epsilon}\left(\left({\bf p}^{\epsilon}\right)^{\star}\right)^{T}{\bf d}\geq{\bf 0}\label{eq:-6}
\end{equation}
for any feasible direction ${\bf d}$, where $\nabla L^{\epsilon}\left(\left({\bf p}^{\epsilon}\right)^{\star}\right)$
is the gradient of the objective function $L^{\epsilon}\left({\bf p}\right)$
at $\left({\bf p}^{\epsilon}\right)^{\star}$.
\begin{prop}
Under Assumption 1, let $\epsilon_{k}$ be a positive sequence with
${\displaystyle \lim_{k\to+\infty}\epsilon_{k}=0}$, then any limit
point ${\bf p}^{\star}$ of the sequence $\left\{ \left({\bf p}^{\epsilon_{k}}\right)^{\star}\right\} $
is a stationary point of problem (\ref{eq:P-sum-of-rank-one}).\end{prop}
\begin{IEEEproof}
The conclusion follows from the continuity of $\nabla L^{\epsilon}\left(\left({\bf p}^{\epsilon}\right)^{\star}\right)$
in $\left({\bf p}^{\epsilon}\right)^{\star}$ and $\epsilon$ under
Assumption 2.
\end{IEEEproof}
In practice, as $\epsilon$ can be chosen as an arbitrarily small
number, directly applying Algorithms \ref{Acc-LIKES}, \ref{Circulant-Embedding}
and \ref{banded-toeplitz} or adapting them to solving the $\epsilon$-approximation
problem would be virtually the same.

\section{Tyler's Estimator with Non-Convex Structure}

In the previous sections we have proposed algorithms for Tyler's estimator
with a general convex structural constraint and discussed in detail
some special cases. For the non-convex structure, the problem is more
difficult to handle. In this section, we are going to introduce two
popular non-convex structures that are tractable by applying the MM
algorithm, namely the spiked covariance structure and the Kronecker
structure.

\subsection{The Spiked Covariance Structure}

The term ``spiked covariance'' was introduced in \cite{johnstone2001distribution}
and refers to the covariance matrix model
\begin{equation}
{\bf R}=\sum_{j=1}^{L}p_{j}{\bf a}_{j}{\bf a}_{j}^{T}+\sigma^{2}{\bf I},\label{eq:spiked-covariance}
\end{equation}
where $L$ is some integer that is less than $K$, and the ${\bf a}_{j}$'s
are unknown orthonormal basis vectors. Note that although (\ref{eq:spiked-covariance})
and (\ref{eq:signal-noise model}) share similar form, they differ
from each other essentially since the ${\bf a}_{j}$'s in (\ref{eq:signal-noise model})
are known and are not necessarily orthogonal. The model is directly
related to principle component analysis, subspace estimation, and
also plays an important role in sensor array applications \cite{Ollila2012Survey,visuri2001subspace}.
This model, referred to as factor model, is also very popular in financial
time series analysis \cite{ruppert2010statistics}.

The constrained optimization problem is formulated as
\begin{equation}
\begin{aligned} & \underset{{\bf R},{\bf a}_{j},{\bf p}\geq{\bf 0},\sigma}{\text{minimize}} &  & \log\det\left({\bf R}\right)+\frac{K}{N}\sum_{i=1}^{N}\log\left({\bf x}_{i}^{T}{\bf R}^{-1}{\bf x}_{i}\right)\\
 & \ \text{subject to} &  & {\bf R}=\sum_{j=1}^{L}p_{j}{\bf a}_{j}{\bf a}_{j}^{T}+\sigma^{2}{\bf I},\\
 &  &  & {\bf A}{\bf A}^{T}={\bf I},
\end{aligned}
\label{eq:P-spiked covariance}
\end{equation}
where ${\bf A}=\left[{\bf a}_{1},\ldots,{\bf a}_{L}\right]$.

Applying the upperbound (\ref{eq:log-upperbound}) for the second
term in the objective function yields the following inner minimization
problem:
\begin{equation}
\begin{aligned} & \underset{{\bf R},{\bf a}_{j},{\bf p}\geq{\bf 0},\sigma}{\text{minimize}} &  & \log\det\left({\bf R}\right)+\frac{K}{N}\sum_{i=1}^{N}\frac{{\bf x}_{i}^{T}{\bf R}^{-1}{\bf x}_{i}}{{\bf x}_{i}^{T}{\bf R}_{t}^{-1}{\bf x}_{i}}\\
 & \ \text{subject to} &  & {\bf R}=\sum_{j=1}^{L}p_{j}{\bf a}_{j}{\bf a}_{j}^{T}+\sigma{\bf I},\\
 &  &  & {\bf A}{\bf A}^{T}={\bf I}.
\end{aligned}
\label{eq:Inner-min-rank-constr.}
\end{equation}
Although the problem is non-convex, a global minimizer can be found
in closed-form as
\begin{eqnarray}
\left(\sigma^{\star}\right)^{2} & = & \frac{1}{K-L}\sum_{j=L+1}^{K}\lambda_{j}\nonumber \\
p_{j}^{\star} & = & \lambda_{j}-\left(\sigma^{\star}\right)^{2}\label{eq:p-opt}\\
{\bf a}_{j}^{\star} & = & {\bf u}_{j},\nonumber
\end{eqnarray}
where $\lambda_{1}\geq\cdots\geq\lambda_{K}$ are the sorted eigenvalues
of matrix ${\displaystyle \frac{K}{N}\sum_{i=1}^{N}\frac{{\bf x}_{i}{\bf x}_{i}^{T}}{{\bf x}_{i}^{T}{\bf R}_{t}^{-1}{\bf x}_{i}}}$
and the ${\bf u}_{j}$'s are the associated eigenvectors \cite{tipping1999probabilistic}.
The algorithm for the spiked covariance structure is summarized in
Algorithm \ref{spike}.

\begin{algorithm}
\caption{\label{spike}Robust covariance estimation under the spiked covariance structure}
\begin{algorithmic}[1]
\State Initialize ${\bf R}_0$ to be an arbitrary feasible positive definite matrix.
\Repeat
\State  ${\bf M}_t = \frac{K}{N}\sum_{i=1}^{N}\frac{{\bf x}_{i}{\bf x}_{i}^{T}}{{\bf x}_{i}^{T}{\bf R}_{t}^{-1}{\bf x}_{i}}$.
\State Eigendecompose ${\bf M}_t$ as ${\bf M}_t=\sum_{j=1}^{L}\lambda_j{\bf u}_j{\bf u}_j^T$, where $\lambda_{1}\geq\cdots\geq\lambda_{K}$.
\State Compute $\sigma^{\star}$, $p_{j}^{\star}$, ${\bf a}_{j}^{\star}$ with \eqref{eq:p-opt}

\State $\tilde{{\bf R}}_{t+1}=\sum_{j=1}^{L}p_{j}^{\star}{\bf a}_{j}^{\star}\left({\bf a}_{j}^{\star}\right)^{T}+\sigma^{\star}{\bf I}.$
\State ${\bf R}_{t+1}=\tilde{{\bf R}}_{t+1}/\text{Tr}\left(\tilde{{\bf R}}_{t+1}\right).$
\State $t \gets t+1$.
\Until{Some convergence criterion is met.}
\end{algorithmic}
\end{algorithm}

As the feasible set is not convex, the convergence statement of the
MM algorithm in \cite{razaviyayn2013unified} needs to be modified
as follows.
\begin{prop}
Any limit point ${\bf R}^{\star}$ generated by the algorithm satisfies
\[
{\rm Tr}\left(\nabla L\left({\bf R}^{\star}\right)^{T}{\bf R}\right)\geq0,\ \forall{\bf R}\in\mathcal{T}_{\mathcal{S}}\left({\bf R}^{\star}\right),
\]
where $\mathcal{T}_{\mathcal{S}}\left({\bf R}^{\star}\right)$ stands
for the tangent cone of $\mathcal{S}$ at ${\bf R}^{\star}$.\end{prop}
\begin{IEEEproof}
The result follows by combining the standard convergence proof of
the MM algorithm \cite{razaviyayn2013unified} and the necessity condition
of ${\bf R}^{\star}$ being the global minimal of $g\left({\bf R}|{\bf R}^{\star}\right)$
over an arbitrary set $\mathcal{S}$ (see Proposition 4.7.1 in \cite{bertsekas2003convex}):
\[
{\rm Tr}\left(\nabla g\left({\bf R}^{\star}|{\bf R}^{\star}\right)^{T}{\bf R}\right)\geq0,\ \forall{\bf R}\in\mathcal{T}_{\mathcal{S}}\left({\bf R}^{\star}\right).
\]

\end{IEEEproof}

\subsection{The Kronecker Structure}

In this subsection we consider the covariance matrix that can be expressed
as the Kronecker product of two matrices, i.e.,
\begin{equation}
{\bf R}={\bf A}\otimes{\bf B},\label{eq:}
\end{equation}
where ${\bf A}\in\mathbb{S}_{+}^{p}$ and ${\bf B}\in\mathbb{S}_{+}^{q}$.

Substituting ${\bf R}={\bf A}\otimes{\bf B}$ into the objective function
yields the equivalent problem:
\begin{equation}
\begin{aligned} & \underset{{\bf A}\succeq{\bf 0},{\bf B}\succeq{\bf 0}}{\text{minimize}} &  & \frac{pq}{N}\sum_{i=1}^{N}\log\text{Tr}\left({\bf A}^{-1}{\bf M}_{i}^{T}{\bf B}^{-1}{\bf M}_{i}\right)\\
 &  &  & +q\log\det\left({\bf A}\right)+p\log\det\left({\bf B}\right)
\end{aligned}
\label{eq:P-kron-constr-equivalent}
\end{equation}
where ${\bf M}_{i}\in\mathbb{R}^{q\times p}$ and ${\bf M}_{i}={\rm vec}\left({\bf x}_{i}\right)$.
Denote the objective function of (\ref{eq:P-kron-constr-equivalent})
as $L\left({\bf A},{\bf B}\right)$.

Note that although the objective function of the equivalent problem
is still non-convex, the constraint set of the equivalent problem
(\ref{eq:P-kron-constr-equivalent}) becomes the Cartesian product
of two convex sets, which is convex.

\subsubsection{Gauss-Seidel}

Since $L\left({\bf R}\right)$ is scale-invariant, we can make the
restriction that ${\rm Tr}\left({\bf A}\right)=1$ and ${\rm Tr}\left({\bf B}\right)=1$
and then problem (\ref{eq:P-kron-constr-equivalent}) can be solved
by updating ${\bf A}$ and ${\bf B}$ alternatively.

Specifically, for fixed ${\bf B}={\bf B}_{t}$, we need to solve the
following problem:
\begin{equation}
\begin{aligned} & \underset{{\bf A}\succeq{\bf 0}}{\text{minimize}} &  & \log\det\left({\bf A}\right)+\frac{p}{N}\sum_{i=1}^{N}\log\text{Tr}\left({\bf A}^{-1}{\bf M}_{i}^{T}{\bf B}_{t}^{-1}{\bf M}_{i}\right)\\
 & \text{subject to} &  & \text{Tr}\left({\bf A}\right)=1.
\end{aligned}
\label{eq:subprob-A}
\end{equation}
Setting the gradient of the objective function to zero yields the
fixed-point equation
\begin{equation}
{\bf A}=\frac{p}{N}\sum_{i=1}^{N}\frac{{\bf M}_{i}^{T}{\bf B}_{t}^{-1}{\bf M}_{i}}{\text{Tr}\left({\bf A}^{-1}{\bf M}_{i}^{T}{\bf B}_{t}^{-1}{\bf M}_{i}\right)}.\label{eq:fixed-point-A}
\end{equation}
As objective function of (\ref{eq:subprob-A}) is essentially the
same as the Tyler's cost function (\ref{eq:Tyler-cost-function}),
an argument similar to Theorem 2.1 in \cite{tyler1987distribution}
reveals that the solution to (\ref{eq:fixed-point-A}) is unique up
to a positive scaling factor, and under Assumption 1, the iteration
\begin{equation}
\begin{aligned}\tilde{{\bf A}} & =\frac{p}{N}\sum_{i=1}^{N}\frac{{\bf M}_{i}^{T}{\bf B}_{t}^{-1}{\bf M}_{i}}{\text{Tr}\left({\bf A}_{r}^{-1}{\bf M}_{i}^{T}{\bf B}_{t}^{-1}{\bf M}_{i}\right)}\\
{\bf A}_{r+1} & =\tilde{{\bf A}}/\text{Tr}\left(\tilde{{\bf A}}\right)
\end{aligned}
\label{eq:update-A}
\end{equation}
converges to the unique global minimum of (\ref{eq:subprob-A}) as
$r\to+\infty$.

Assign ${\bf A}_{t+1}=\underset{r\to+\infty}{\lim}{\bf A}_{r}$, similarly
we have the fixed-point iteration for ${\bf B}$ as
\begin{equation}
\begin{aligned}\tilde{{\bf B}} & =\frac{q}{N}\sum_{i=1}^{N}\frac{{\bf M}_{i}{\bf A}_{t+1}^{-1}{\bf M}_{i}^{T}}{\text{Tr}\left({\bf A}_{t+1}^{-1}{\bf M}_{i}^{T}{\bf B}_{r}{\bf M}_{i}\right)}\\
{\bf B}_{r+1} & =\tilde{{\bf B}}/\text{Tr}\left(\tilde{{\bf B}}\right),
\end{aligned}
\label{eq:update-B}
\end{equation}
and ${\bf B}_{t+1}=\underset{r\to+\infty}{\lim}{\bf B}^{r}$.
\begin{prop}
Under Assumption 1, every limit point of the sequence $\left\{ \left({\bf A}_{t},{\bf B}_{t}\right)\right\} $
generated by Algorithm \ref{Gauss-Seidel} is a stationary point of
(\ref{eq:P-kron-constr-equivalent}).\end{prop}
\begin{IEEEproof}
Application of Proposition 2.7.1 in \cite{Bertsekas1999}.
\end{IEEEproof}
\begin{algorithm}
\caption{\label{Gauss-Seidel}Robust covariance estimation under the Kronecker structure (Gauss-Seidel)}
\begin{algorithmic}[1]
\State Initialize ${\bf A}_0$ and ${\bf B}_0$ to be arbitrary positive definite matrices of size $p\times p$ and $q\times q$, respectively.
\Repeat
\State Update $\bf A$ with \eqref{eq:update-A}.
\State Update $\bf B$ with \eqref{eq:update-B}.
\State $t \gets t+1$.
\Until{Some convergence criterion is met.}
\end{algorithmic}
\end{algorithm}

\subsubsection{Block Majorization Minimization}

A stationary point of $L\left({\bf A},{\bf B}\right)$ can also be
found by block majorization minimization algorithm (Block MM).

By Proposition \ref{prop:full-mm-bound}, with the value of ${\bf B}_{t}$
fixed to be ${\bf B}_{t}$, a convex upperbound of $L\left({\bf A},{\bf B}\right)$
on $\mathbb{S}_{+}^{p}$ at point ${\bf A}_{t}$ (ignoring a constant
term and up to a scale factor of $q$) can be found as
\begin{equation}
g\left({\bf A}|{\bf A}_{t},{\bf B}_{t}\right)={\rm Tr}\left({\bf A}_{t}^{-1}{\bf A}\right)+\frac{p}{N}\sum_{i=1}^{N}\frac{\text{Tr}\left({\bf A}^{-1}{\bf M}_{i}^{T}{\bf B}_{t}^{-1}{\bf M}_{i}\right)}{\text{Tr}\left({\bf A}_{t}^{-1}{\bf M}_{i}^{T}{\bf B}_{t}^{-1}{\bf M}_{i}\right)}.\label{eq:surrogate function-kron-1}
\end{equation}

\begin{lem}
\label{lem:M>0-1}Under Assumption 1, for any ${\bf A}_{t},{\bf B}_{t}\succ{\bf 0}$,
the matrix
\[
{\bf M}\left({\bf A}_{t},{\bf B}_{t}\right)=\frac{p}{N}\sum_{i=1}^{N}\frac{{\bf M}_{i}^{T}{\bf B}_{t}^{-1}{\bf M}_{i}}{\text{Tr}\left({\bf A}_{t}^{-1}{\bf M}_{i}^{T}{\bf B}_{t}^{-1}{\bf M}_{i}\right)}
\]
is nonsingular.\end{lem}
\begin{IEEEproof}
At $\left({\bf A}_{t},{\bf B}_{t}\right)$ (ignoring a constant term
and up to a scale factor of $q$) the function $L\left({\bf A},{\bf B}_{t}\right)$
can be upperbounded by
\begin{equation}
\tilde{g}\left({\bf A}|{\bf A}_{t},{\bf B}_{t}\right)=\log\det\left({\bf A}\right)+\text{Tr}\left({\bf A}^{-1}{\bf M}\left({\bf A}_{t},{\bf B}_{t}\right)\right).\label{eq:Gaussianize A-1}
\end{equation}
If ${\bf M}\left({\bf A}_{t},{\bf B}_{t}\right)$ is singular, we
can eigendecompose ${\bf M}\left({\bf A}_{t},{\bf B}_{t}\right)$
as ${\bf M}\left({\bf A}_{t},{\bf B}_{t}\right)={\bf U}{\rm diag}\left(\lambda_{1},\ldots,\lambda_{p}\right){\bf {\bf U}}^{T}$
with $\lambda_{1}=0$, and set ${\bf A}^{-1}={\bf U}{\rm diag}\left(\sigma_{1},\ldots,\sigma_{p}\right){\bf {\bf U}}^{T}$.

Letting $\sigma_{1}\to0$ would lead to $\tilde{g}\left({\bf A}|{\bf A}_{t},{\bf B}_{t}\right)$
unbounded below, which implies $L\left({\bf A},{\bf B}_{t}\right)$
is also unbounded below and contradicts Assumption 1.
\end{IEEEproof}
An immediate implication of Lemma \ref{lem:M>0-1} is that $g\left({\bf A}|{\bf A}_{t},{\bf B}_{t}\right)$
is strictly convex on $\mathbb{S}_{++}^{p}$ and has a unique closed-form
minimizer given by
\begin{equation}
{\bf A}_{t+1}={\bf A}_{t}^{1/2}\left({\bf A}_{t}^{-1/2}{\bf M}{\bf A}_{t}^{-1/2}\right)^{1/2}{\bf A}_{t}^{1/2},\label{eq:karcher-mean-A}
\end{equation}
where ${\displaystyle {\bf M}=\frac{p}{N}\sum_{i=1}^{N}\frac{{\bf M}_{i}^{T}{\bf B}_{t}^{-1}{\bf M}_{i}}{\text{Tr}\left({\bf A}_{t}^{-1}{\bf M}_{i}^{T}{\bf B}_{t}^{-1}{\bf M}_{i}\right)}}.$

Symmetrically, we have the the update for ${\bf B}$ given by
\begin{equation}
{\bf B}_{t+1}={\bf B}_{t}^{1/2}\left({\bf B}_{t}^{-1/2}{\bf M}{\bf B}_{t}^{-1/2}\right)^{1/2}{\bf B}_{t}^{1/2},\label{eq:karcher-mean-B}
\end{equation}
where ${\displaystyle {\bf M}=\frac{q}{N}\sum_{i=1}^{N}\frac{{\bf M}_{i}{\bf A}_{t+1}^{-1}{\bf M}_{i}^{T}}{\text{Tr}\left({\bf A}_{t+1}^{-1}{\bf M}_{i}^{T}{\bf B}_{t}^{-1}{\bf M}_{i}\right)}}$.
\begin{prop}
Under Assumption 1, every limit point of the pair generated by Algorithm
\ref{BSUM} is a stationary point of the problem (\ref{eq:P-kron-constr-equivalent}).\end{prop}
\begin{IEEEproof}
Application of Theorem 2 (a) in \cite{razaviyayn2013unified}.
\end{IEEEproof}
Compared to Algorithm \ref{Gauss-Seidel}, which is a double loop
algorithm, Algorithm \ref{BSUM} only performs a single loop iteration.

Note that with the surrogate function of the form (\ref{eq:surrogate function-kron-1}),
we can easily impose additional convex structures on ${\bf A}$ and
${\bf B}$, and the update is found by solving the convex problem:
\begin{equation}
\begin{aligned}{\bf A}_{t+1} & =\arg\min_{{\bf A}\in\mathcal{A}}g\left({\bf A}|{\bf A}_{t},{\bf B}_{t}\right),\\
{\bf B}_{t+1} & =\arg\min_{{\bf B}\in\mathcal{B}}g\left({\bf B}|{\bf A}_{t+1},{\bf B}_{t}\right),
\end{aligned}
\label{eq:-7}
\end{equation}
with $\mathcal{A}$ and $\mathcal{B}$ being the convex structural
constraint sets.

\begin{algorithm}
\caption{\label{BSUM}Robust covariance estimation under the Kronecker structure (Block Majorization Minimization)}
\begin{algorithmic}[1]
\State Initialize ${\bf A}_0$ and ${\bf B}_0$ to be arbitrary positive definite matrices of size $p\times p$ and $q\times q$,  respectively.
\Repeat
\State Update $\bf A$ with \eqref{eq:karcher-mean-A}.
\State Update $\bf B$ with \eqref{eq:karcher-mean-B}.
\State $t \gets t+1$.
\Until{Some convergence criterion is met.}
\end{algorithmic}
\end{algorithm}

\section{Numerical Results}

In this section, we present numerical results that demonstrate the
effect of imposing structure on the covariance estimator on reducing
estimation error, and provide a comparison of the proposed estimator
with some state-of-the-art estimators. The estimation error is evaluated
by the normalized mean-square error, namely
\begin{equation}
\textrm{NMSE}\left(\hat{{\bf R}}\right)=\frac{E\left\Vert \hat{{\bf R}}-{\bf R}_{0}\right\Vert _{F}^{2}}{\left\Vert {\bf R}_{0}\right\Vert _{F}^{2}},\label{eq:NMSE}
\end{equation}
where all of the matrices are normalized by their trace. The expected
value is approximated by 100 Monte Carlo simulations. In the following,
we mainly compare the performance of four estimators, namely, the
SCM, unconstrained Tyler's estimator (fixed-point equation of (\ref{eq:Tyler-estimator})),
COCA (solution to (\ref{eq:COCA})), and the proposed structure constrained
Tyler's estimator. The samples in all of the simulations of this section,
if not otherwise specified, are i.i.d. following ${\bf x}_{i}\sim\sqrt{\tau}{\bf u}$,
where $\tau\sim\chi^{2}$ and ${\bf u}\sim\mathcal{N}\left({\bf 0},{\bf R}_{0}\right)$.
The dimension $K$ is set to be $15$.

\subsection{Toeplitz Structure}

\begin{figure}
\begin{centering}
\includegraphics[scale=0.5]{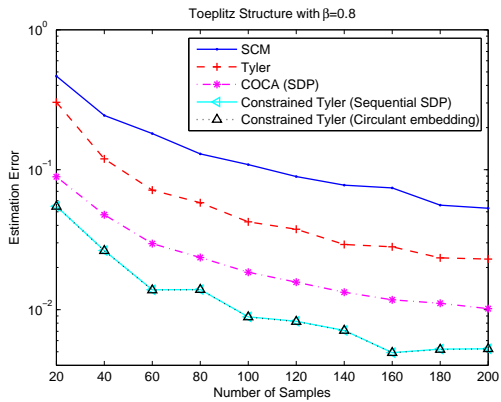}
\par\end{centering}

\protect\caption{\label{fig:Toeplitz}The estimation error (NMSE) of different estimators
under the Toeplitz structure of the form (\ref{eq:AR-Toeplitz}).}

\end{figure}

\begin{figure}
\begin{centering}
\includegraphics[scale=0.5]{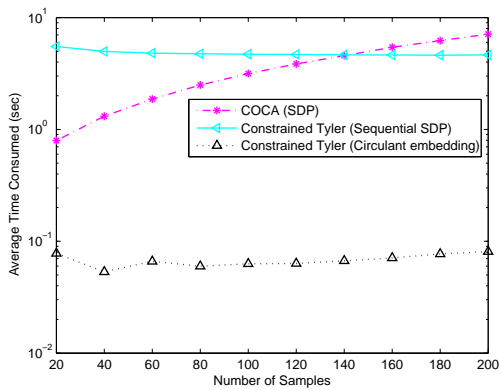}
\par\end{centering}

\protect\caption{\label{fig:time-toeplitz}Average time (in seconds) consumed by COCA
and the constrained Tyler's estimator via sequential SDP (Algorithm
1) and circulant embedding (Algorithm 2).}
\end{figure}

In this simulation, ${\bf R}_{0}$ is set to be a Toeplitz matrix.
The parameter ${\bf R}_{0}$ is set to be ${\bf R}\left(\beta\right)$,
whose $ij$-th entry is of the form
\begin{equation}
\left({\bf R}\left(\beta\right)\right)_{ij}=\beta^{\left|i-j\right|}.\label{eq:AR-Toeplitz}
\end{equation}
Fig. \ref{fig:Toeplitz} shows the NMSE of the estimators with $\beta=0.8$
. The result indicates that the structure constrained Tyler's estimator
achieves the smallest estimation error. In addition, we see that although
the circulant embedding algorithm (Algorithm \ref{Acc-LIKES}) with
$L=2K-1$ approximately solves the Toeplitz structure constrained
problem, it achieves virtually the same estimation error as imposing
the Toeplitz structure and solving the problem via the sequential
SDP algorithm (Algorithm \ref{SCA}). However, the computational cost
of circulant embedding is much lower than that of sequential SDP and
COCA, as shown in the average time cost plotted in Fig. \ref{fig:time-toeplitz}.

\subsection{Banded Toeplitz Structure}

\begin{figure}
\begin{centering}
\includegraphics[scale=0.5]{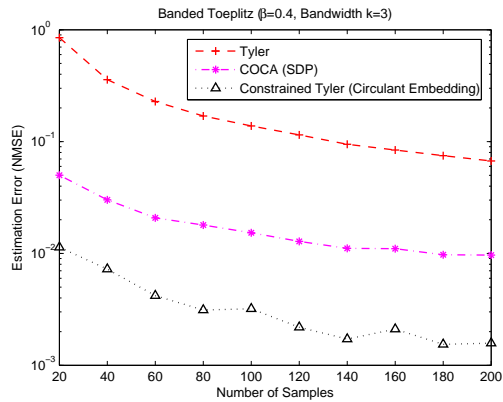}\protect\caption{\label{fig:banded Toeplitz}The estimation error (NMSE) of different
estimators under the banded Toeplitz structure.}

\par\end{centering}

\end{figure}

\begin{figure}
\begin{centering}
\includegraphics[scale=0.5]{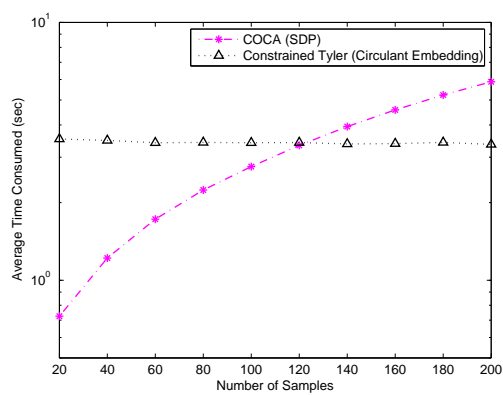}
\par\end{centering}

\begin{centering}
\protect\caption{\label{fig:Average-time-banded-Toeplitz}Average time (in seconds)
consumed by COCA and constrained Tyler's estimator.}

\par\end{centering}

\end{figure}

Next we investigate the case that ${\bf R}_{0}$ is a $k$-banded
Toeplitz matrix $B_{k}\left({\bf R}_{0}\right)$, where $B_{k}\left({\bf R}_{0}\right)$
defines a matrix with the $ij$-th entry equals to that of ${\bf R}_{0}$
if $\left|i-j\right|\leq k$, and equals zero otherwise. ${\bf R}_{0}={\bf R}\left(0.4\right)$
and the bandwidth $k$ is chosen to be $3$. The NMSE is plotted in
Fig. \ref{fig:banded Toeplitz}, where the constrained Tyler's estimator
achieves the smallest estimation error. Fig. \ref{fig:Average-time-banded-Toeplitz}
plots the average time consumed by COCA and the constrained Tyler's
estimator. As the number of semidefinite constraints that COCA has
is proportional to $N$, the time consumption is approximately linearly
increasing in $N$, while the time cost by the algorithm for the constrained
Tyler's estimator remains roughly the same as $N$ grows.\textcolor{black}{{}
When $N$ is small, the algorithm for COCA runs faster than ours since
the scale of the SDP that COCA solves is small.} In the regime that
$N$ is large, the computational cost of COCA increases, as reflected
both in the time and the memory required to run the algorithm.

\begin{figure}
\begin{centering}
\includegraphics[scale=0.5]{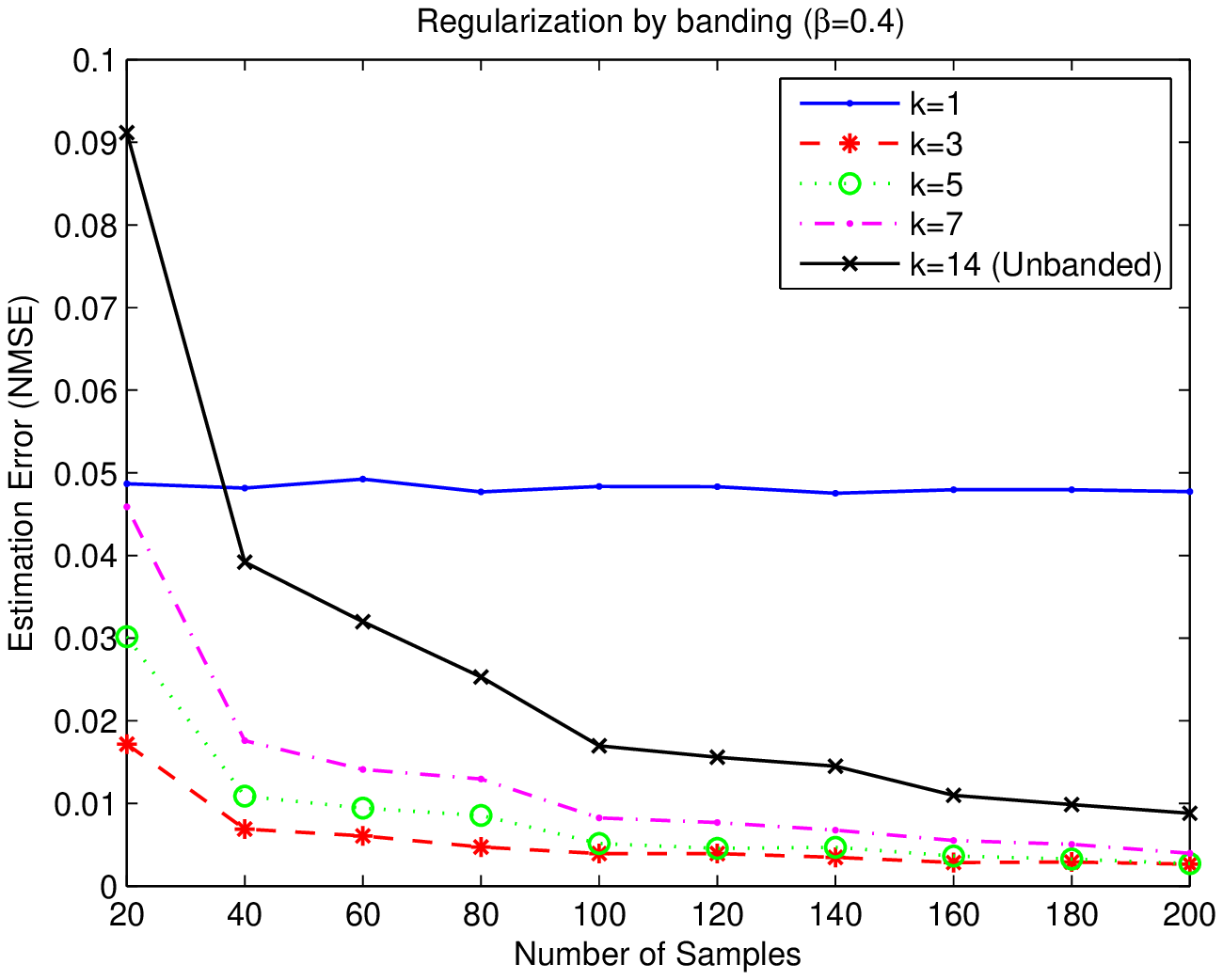}
\par\end{centering}

\begin{centering}
\includegraphics[scale=0.5]{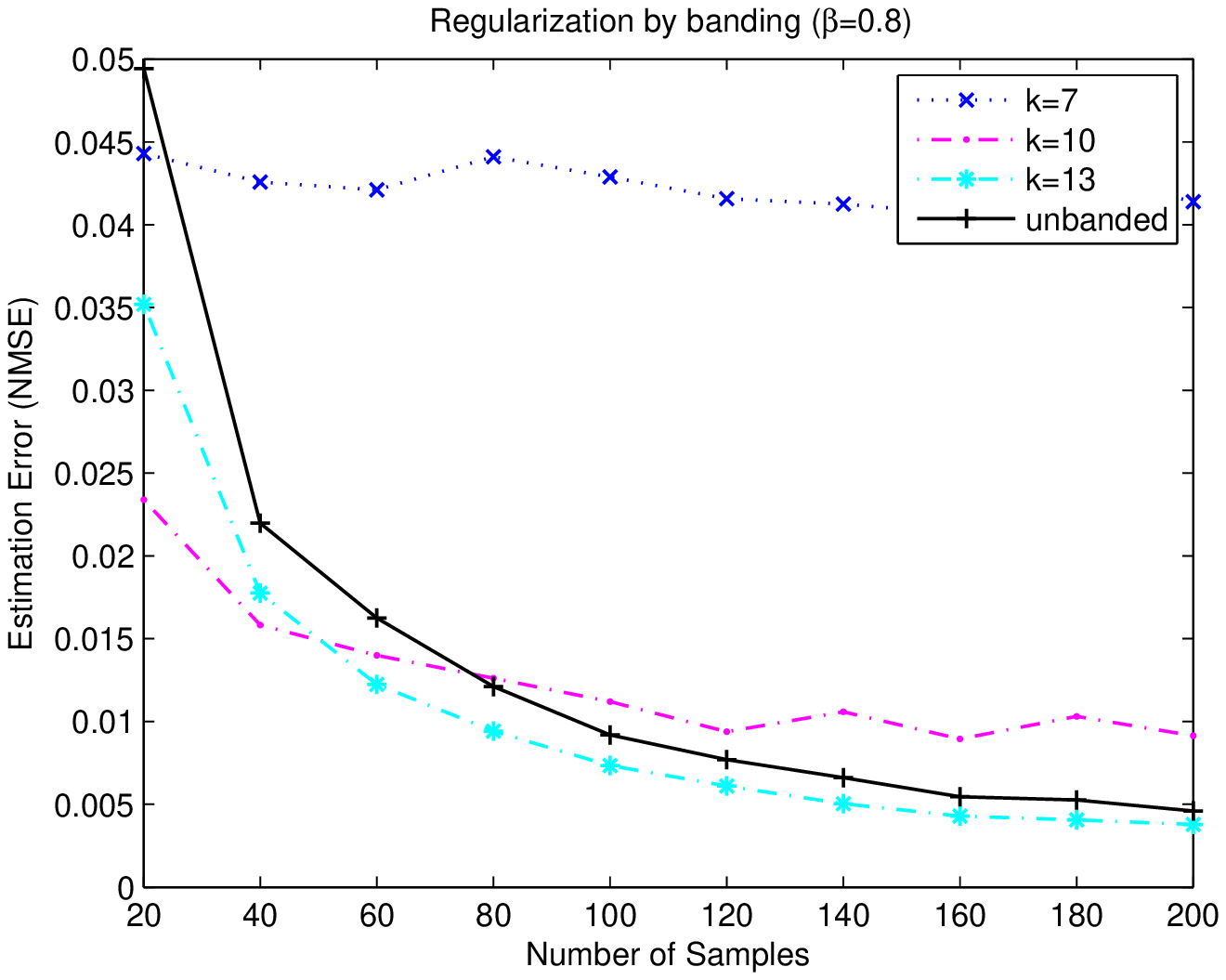}
\par\end{centering}

\protect\caption{\label{fig:regularize by banding (0.4)}NMSE of the regularized Tyler's
estimator by imposing the banded Toeplitz structure of different bandwidth
$k$ when ${\bf R}_{0}={\bf R}\left(0.4\right)$ and ${\bf R}_{0}={\bf R}\left(0.8\right)$.}

\end{figure}

In the third simulation, we consider ${\bf R}_{0}$ being a non-banded
Toeplitz matrix with the property that $\left({\bf R}_{0}\right)_{ij}$
decays rapidly as $\left|i-j\right|$ increases. We investigate the
cases of ${\bf R}_{0}={\bf R}\left(0.4\right)$ (fast decay) and ${\bf R}_{0}={\bf R}\left(0.8\right)$
(slow decay) and impose a banded Toeplitz structure on the Tyler's
estimator with a varying bandwidth $k$ to regularize the estimator.
Fig. \ref{fig:regularize by banding (0.4)} shows that the smallest
error is obtained when $k=3$ in the $\beta=0.4$ case, and when $k=13$
in the $\beta=0.8$ case. In either case, with the right choice of
bandwidth $k$, the regularized estimator outperforms the unbanded
one when the number of samples is relatively small compared to the
dimension of the covariance matrix to be estimated.

\subsection{Direction of Arrival Estimation }

In this subsection, we examine the robustness of the proposed estimator
in the context of the direction of arrival estimation problem with
the following signal model:

\[
{\bf x}\left(t\right)={\bf A}{\bf s}\left(t\right)+{\bf n}\left(t\right),
\]
where ${\bf A}=\left[{\bf a}\left(\theta_{1}\right),\ldots,{\bf a}\left(\theta_{L}\right)\right]$
is the steering matrix and ${\bf n}$ is zero mean additive noise.
We study the simple case of an ideal uniform linear array (ULA) with
half-wavelength inter-element spacing, where
\[
{\bf a}\left(\theta\right)=\left[1,e^{-j\pi\sin\left(\theta\right)},\ldots,e^{-j\pi\left(K-1\right)\sin\left(\theta\right)}\right]^{T}.
\]
Assuming that the signal ${\bf s}\left(t\right)$ is a wide-sense
stationary random process with zero mean, the covariance of ${\bf x}\left(t\right)$
is
\[
{\bf R}={\bf A}{\rm Cov}\left({\bf s}\right){\bf A}^{H}+{\rm Cov}\left({\bf n}\right).
\]
Further assume that the signals arriving from different directions
are uncorrelated and that the noise is spatially white, i.e., ${\rm Cov}\left({\bf s}\right)={\rm diag}\left(p_{1},\ldots,p_{L}\right)$
and ${\rm Cov}\left({\bf n}\right)=\sigma^{2}{\bf I}$, the covariance
model simplifies to be
\[
{\bf R}=\sum_{j=1}^{L}p_{j}{\bf a}\left(\theta_{j}\right){\bf a}\left(\theta_{j}\right)^{H}+\sigma^{2}{\bf I}.
\]
We assume that the number of signals $L$ is known in prior.

In our simulation, 5 random signals are assumed arriving from directions
$-10\textdegree$, $10\textdegree$, $15\textdegree$, $35\textdegree$,
$40\textdegree$ with equal power $p=1$ and the noise power is set
to be $\sigma^{2}=0.1$. The received signal is assumed to be elliptically
distributed. The number of sensors is $K=15$.

\begin{figure}
\begin{centering}
\includegraphics[scale=0.5]{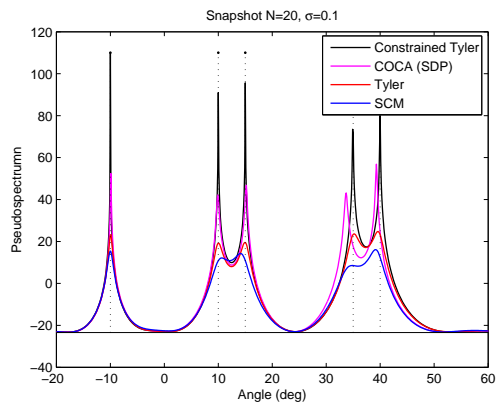}
\par\end{centering}

\protect\caption{\label{fig:DOA-MUSIC}Arrival angle estimated by MUSIC with different
covariance estimators. }

\end{figure}

We first estimate ${\bf R}$ and then apply the MUSIC algorithm to
estimate the arriving angles. The performance of SCM, Tyler's estimator,
COCA and the constrained Tyler's estimator are compared. For the latter
two estimators, ${\bf A}$ is constructed with the $\theta_{l}$'s
uniformly located on the interval $\left[-\pi/2,\pi/2\right]$ with
a stepsize of $5\textdegree$. Fig. \ref{fig:DOA-MUSIC} shows the
estimated arrival direction using different estimators with the number
of snapshots $N=20$, and only the constrained Tyler's estimator correctly
recovers all of the arriving angles.

\begin{figure}
\begin{centering}
\subfloat[\label{fig:NMSE}NMSE]{\begin{centering}
\includegraphics[scale=0.5]{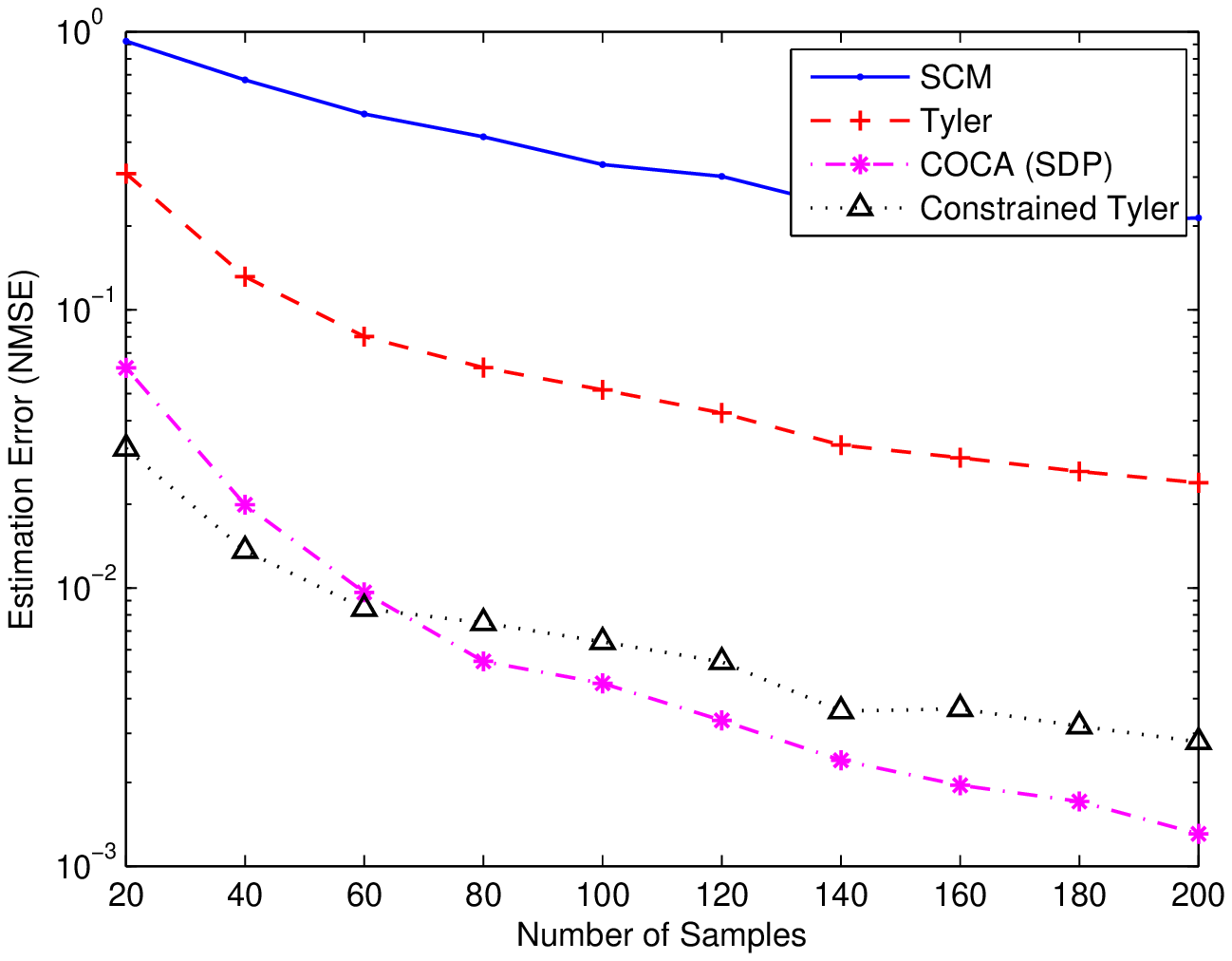}
\par\end{centering}

}
\par\end{centering}

\begin{centering}
\subfloat[\label{fig:Subspace-error}Subspace error]{\begin{centering}
\includegraphics[scale=0.5]{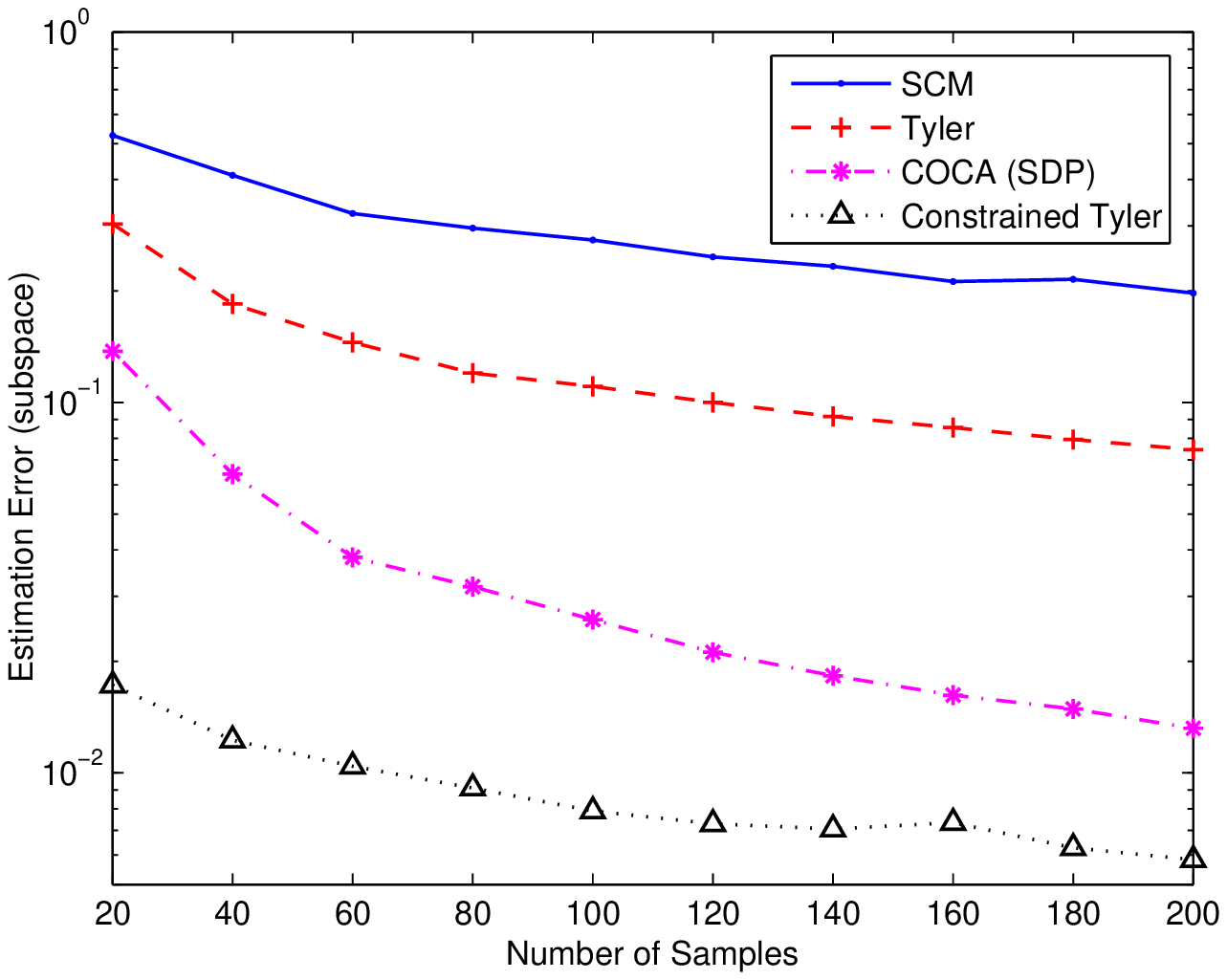}
\par\end{centering}

}
\par\end{centering}

\protect\caption{\label{fig:Avg-error} The estimation error of different estimators
under the DOA structure: (a) NMSE, (b) estimation error of the noise
subspace given by different estimators evaluated by (\ref{eq:subspace error}).}

\end{figure}

Fig. \ref{fig:Avg-error} shows the performance of different estimators
in terms of NMSE and the estimation error of noise subspace evaluated
by
\begin{equation}
\left\Vert \hat{{\bf E}}_{c}\hat{{\bf E}}_{c}^{H}-{\bf E}_{c}{\bf E}_{c}^{H}\right\Vert _{F},\label{eq:subspace error}
\end{equation}
with $N$ varying from 20 to 200, where ${\bf E}_{c}$ denotes the
noise subspace and $\hat{{\bf E}}_{c}$ denotes its estimate.\textcolor{red}{{}
}\textcolor{black}{Fig. 7 (a) reveals that the constrained Tyler's
estimator achieves the smallest NMSE when $N$ is small, while COCA
performs better when $N$ is large.}\textcolor{red}{{} }However, Fig.
7 (b) indicates that the constrained Tyler's estimator can estimate
the noise subspace more accurately for all values of $N$, which is
beneficial for algorithms that are based on $\hat{{\bf E}}_{c}$ such
as MUSIC.

\begin{figure}
\begin{centering}
\includegraphics[scale=0.5]{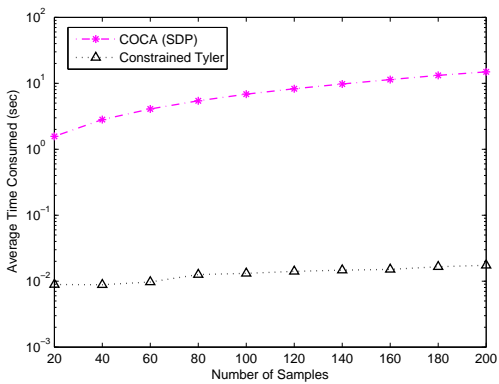}
\par\end{centering}

\protect\caption{\label{fig:Avg-exe-time}Average time (in seconds) consumed per data
set by COCA and constrained Tyler's estimator.}
\end{figure}

The average time cost by COCA and the constrained Tyler's estimator
is plotted in Fig. \ref{fig:Avg-exe-time}. It can be seen that the
proposed method is much faster than COCA. In addition, unlike COCA,
the consumed time of our algorithm is not sensitive to the number
of samples $N$.

\subsection{Spiked Covariance Structure}

\begin{figure}
\begin{centering}
\subfloat[NMSE]{\begin{centering}
\includegraphics[scale=0.5]{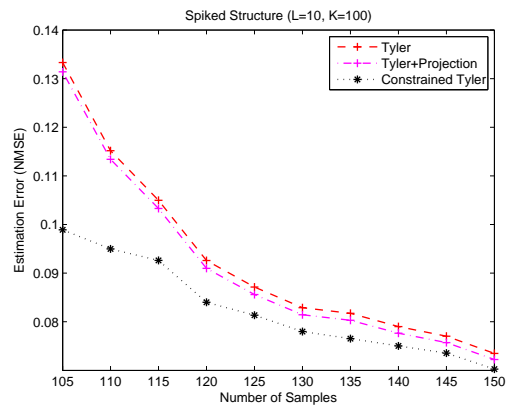}
\par\end{centering}

}
\par\end{centering}

\begin{centering}
\subfloat[Subspace error]{\begin{centering}
\includegraphics[scale=0.5]{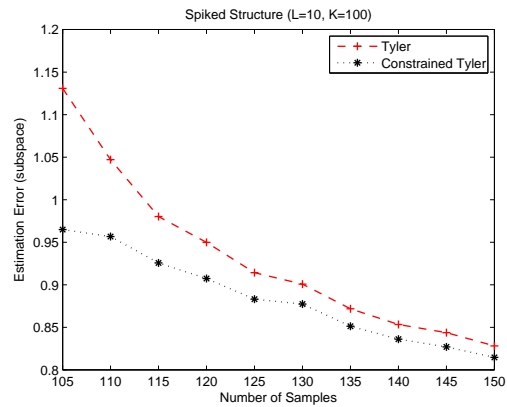}
\par\end{centering}

}
\par\end{centering}

\protect\caption{\label{fig:Spiked-cov-est}The estimation error of different estimators
under the spiked covariance structure: (a) NMSE, (b) estimation error
of the noise subspace given by different estimators evaluated by (\ref{eq:subspace error}).}

\end{figure}

We construct the true covariance ${\bf R}_{0}$ by the following model:
\[
{\bf R}_{0}=\sum_{j=1}^{L}p_{j}{\bf a}_{j}{\bf a}_{j}{}^{H}+\sigma^{2}{\bf I},
\]
where the ${\bf a}_{j}$'s are randomly generated orthonormal basis
and the $p_{j}$'s are randomly generated corresponding eigenvectors
uniformly distributed in $\left[0.01,1\right]$. $\sigma^{2}$ is
set to be $0.01$. The number of spikes $L=10$ is assumed to be known
in prior. The matrix dimension is fixed to be $K=100$, and the number
of samples is varied from $N=105$ to $N=150$. As COCA applies only
for convex structural set and cannot be used here, we replace it by
the projected Tyler's estimator, which is a two step procedure that
first obtains the Tyler's estimator and then performs projection according
to (\ref{eq:p-opt}). Fig. \ref{fig:Spiked-cov-est} shows that imposing
the spiked structure helps in reducing the NMSE and subspace estimation
error measured by (\ref{eq:subspace error}).

\subsection{Kronecker Structure}

\begin{figure}
\begin{centering}
\includegraphics[scale=0.55]{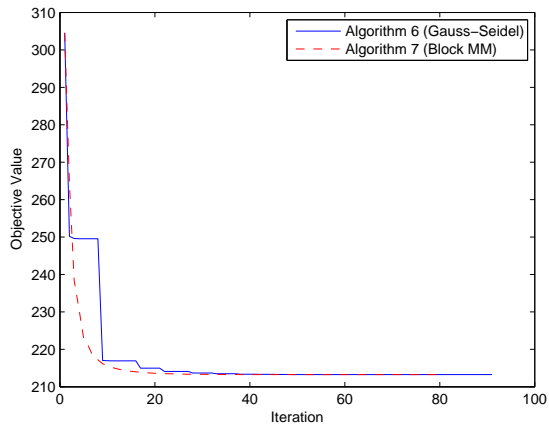}
\par\end{centering}

\protect\caption{\label{fig:Convergence-Comparison-Kron}Convergence Comparison of
Algorithm \ref{Gauss-Seidel} and \ref{BSUM} under the Kronecker
structure.}
\end{figure}

The parameters are set to be ${\bf A}_{0}={\bf I}$ , ${\bf B}_{0}={\bf R}\left(0.8\right)$,
$p=10$, $q=8$, in the simulations. We first plot the convergence
curve of Algorithms \ref{Gauss-Seidel} and \ref{BSUM} with the number
of samples $N=4$ in Fig. \ref{fig:Convergence-Comparison-Kron}.
The two algorithms converges in roughly the same number of iterations,
and the objective value corresponds to Algorithm \ref{BSUM} (block
MM) decreases more smoothly than Algorithm \ref{Gauss-Seidel} (Gauss-Seidel),
as the latter is a double loop algorithm while the former is a single
loop algorithm.

\begin{figure}
\begin{centering}
\includegraphics[scale=0.55]{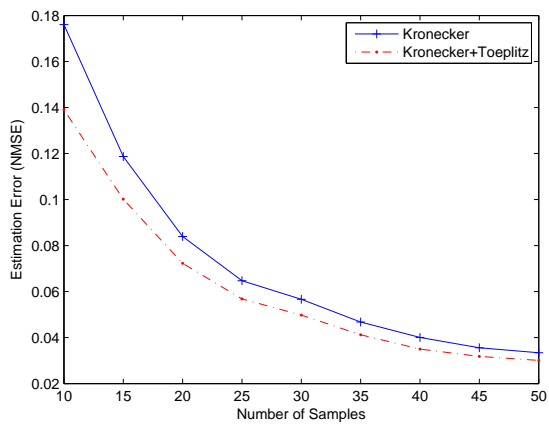}
\par\end{centering}

\protect\caption{\label{fig:NMSE-kron} NMSE of Tyler's estimator with a Kronecker
structural constraint versus that with both a Kronecker and a Toeplitz
structural constraint.}
\end{figure}

Fig. \ref{fig:NMSE-kron} plots the NMSE of Tyler's estimator with
a Kronecker constraint on ${\bf R}$ and that with both a Kronecker
constraint on ${\bf R}$ and a Toeplitz constraint on ${\bf B}$.
We can see that further imposing a Toeplitz structure on ${\bf B}$
helps in reducing the estimation error.

\section{Conclusion}

In this paper, we have discussed the problem of robustly estimating
the covariance matrix with a prior structure information. The problem
has been formulated as minimizing the negative log-likelihood function
of the angular central Gaussian distribution subject to the prior
structural constraint. For the general convex constraint, we have
proposed a sequential convex programming algorithm based on the majorization
minimization framework. The algorithm has been particularized with
higher computational efficiency for several specific structures that
are widely considered in the signal processing community. The spiked
covariance model and the Kronecker structure, although belonging to
the non-convex constraint, are also discussed and shown to be computationally
tractable. The proposed estimator has been shown \textcolor{black}{outperform
the state-of-the-art methods in the numerical section}.

\bibliographystyle{IEEEtran}
\bibliography{refs}

\end{document}